\documentclass[british,english]{article}
\usepackage[T1]{fontenc}
\usepackage[latin9]{inputenc}
\usepackage{geometry}
\usepackage{color}
\usepackage{verbatim}
\usepackage{textcomp}
\usepackage{amsmath}
\usepackage{amssymb}
\usepackage{graphicx}

\makeatletter

\providecommand{\tabularnewline}{\\}

\newcommand{\lyxaddress}[1]{
\par {\raggedright #1
\vspace{1.4em}
\noindent\par}
}
\newenvironment{lyxlist}[1]
{\begin{list}{}
{\settowidth{\labelwidth}{#1}
 \setlength{\leftmargin}{\labelwidth}
 \addtolength{\leftmargin}{\labelsep}
 }}
{\end{list}}
\newenvironment{lyxcode}
{\par\begin{list}{}{
\setlength{\rightmargin}{\leftmargin}
\setlength{\listparindent}{0pt}
\raggedright
\setlength{\itemsep}{0pt}
\setlength{\parsep}{0pt}
\normalfont\ttfamily}%
 \item[]}
{\end{list}}

\makeatother

\usepackage{babel}
\newcommand{\Ext}{\mathrm{Ext}}
\newcommand{\Bd}{\mathrm{Bd}}
\newcommand{\Cl}{\mathrm{Cl}}
\newcommand{\E}{_{\mathrm{\scriptstyle{E}}}}
\newcommand{\NE}{_{\mathrm{\scriptstyle{NE}}}}
\newcommand{\Ob}{^{\mathrm{O}}}

\begin{document}

\title{A topological approach to the problem of \\emergence in complex
systems}

\author{Alberto Pascual-García\\}

\maketitle

\lyxaddress{\begin{center}
\textcolor{black}{\footnotesize{}Centro de Biología Molecular ``Severo
Ochoa'' (CSIC-UAM).\\ Nicolás Cabrera 1, campus-UAM, E-28049. Madrid
(Spain). \\}\textbf{\textcolor{black}{\footnotesize{}Present Address:}}\textcolor{black}{\footnotesize{}
Department of Life Sciences \\ Imperial College London, Ascot, Berkshire,
UK \\ }\textbf{\footnotesize{}Email. }\footnotesize{}alberto.pascual.garcia@gmail.com 
\par\end{center}}
\begin{abstract}
Emergent patterns in complex systems are related to many intriguing
phenomena in modern science and philosophy. Several conceptions such
as weak, strong and robust emergence have been proposed to emphasize
different epistemological and ontological aspects of the problem.
One of the most important concerns is whether emergence is an intrinsic
property of the reality we observe, or it is rather a consequence
of epistemological limitations. To elucidate this question, we propose
a novel approximation through constructive topology, a framework that
allow us to map the space of observed objects (ontology) with the
knowledge subject conceptual apparatus (epistemology). Focusing in
a particular type of emergent processes, namely those accessible through
experiments and from which we have still no clue on the mechanistic
processes yielding its formation, we analyse
how a knowledge subject would build a conceptual explanatory framework,
what we will call an arithmomorphic scheme. Working on these systems,
we identify concept disjunction as a critical logical operation needed
to identify the system's constraints. Next, focusing on a three-bits
synthetic system, we show how the number and scope of the constraints
hinder the development of an arithmomorphic scheme. Interestingly,
we observe that our framework is unable to identify global constraints,
clearly linking the epistemological limits of the framework with an
ontological feature of the system. This allows us to propose a definition
of emergence strength which we make compatible with the scientific
method through the active intervention of the observer on the system,
following the spirit of Granger causality. We think that this definition
reconciles previous attempts to classify emergent processes, at least
for the specific kind we discuss here. The paper finishes discussing
the relevance of global constraints in biological systems, understood
as a downward causal influence exerted by natural selection. In summary,
we think that our approach provides a meeting point for previous efforts
on this topic, and we expect that it will stimulate further research
in both scientific and philosophical communities.
\end{abstract}

\section{Introduction}
\begin{quotation}
\emph{``What urges you on and arouses your,
you wisest of men, do you call it \textquotedbl{}will to truth\textquotedbl{}?
Will to the conceivability
of all being: that is what I call your will! You first want to make
all being conceivable: for, with a healthy mistrust, you doubt whether
it is in fact conceivable. But it must bend and accommodate itself
to you! Thus will your will have it. It must become smooth and subject
to the mind as the mind's mirror and reflection.}`` 

Friedrich Nietzsche \cite{nietzsche_Zaratustra_2014}
\end{quotation}
Scientific modelling
is probably one of the best examples of a human activity fitting the
words of Zarathustra: it requires the generation of conceptual representations
for processes which depend, many times, on uncomfortable features
such as measurement inaccuracy, constituents interdependence or dynamics.
We attempt to incorporate these representations within a mathematical
or computational framework, which is nothing but a comfortable place
where we reaffirm our confidence in the acquired knowledge. Building
a formal framework provides a favourable environment for reaching new
analytical and computational results, thus accelerating the outcome
of new predictions that can be firmly settled within the scientific
knowledge after hypothesis testing.

One of the most interesting challenges in scientific modelling
relies on complex systems, which are systems composed
by a large number of entities driven by non-linear interactions between
their components and with the environment. In particular, complex
systems may lead to the observation of one of the most controversial
phenomena in modern sciences: emergent behaviours.
A good intuition for emergence arises when we observe in a complex
system that it ``begins to exhibit genuinely novel properties that
{[}at a first sight{]}\footnote{The statement in brackets is ours because, although we think that
this definition provides a good intuition on emergent processes, we
agree with Mitchell in that it is simplistic \cite{mitchell_EmergAntiKim_2012}.
We rather adhere to the view of Francis Crick when he explains that
``The scientific meaning of emergent, or at least the one I use,
assumes that, while the whole may not be the simple sum of its separate
parts, its behaviour can, at least in principle, be understood from
the nature and behaviour of its parts plus the knowledge of how all
these parts interact'' (reproduced from \cite{corning_reemergence_2012}
referring to \cite{crick_astonishingHyp_1994} p.11)}are irreducible to, and neither predictable nor explainable in terms
of, the properties of their constituents''\cite{kim_MakeSenseEmerg_1999}.
Among this kind of collective behaviours we find phenomena such as
magnetism, patterns observed in dissipative systems like hurricanes
or convection cells or, in biological systems, patterns on animal
skins or flocking behaviour. Looking at these examples it seems that
the difficulty relies in an apparent discontinuity between emergent
properties and their microscopic description. Since a basic tenet
in the scientific method is that macroscopic properties are the consequence
of the lower level constituents a critical question arises here \cite{brenner_SeqsConseqs_2010}:
how is it possible to obtain a satisfactory conceptual representation
of emergent macroscopic behaviours when the definition of emergence
apparently implies a discontinuity between the microscopic and the
macroscopic representation? 

Explaining the origin of this discontinuity has led to the famous
controversy between vitalist and reductionists positioning \cite{emmeche_explaining_1997,corning_reemergence_2012}.
We are not interested in entering into this debate, because we do
not aim to understand which are the mechanisms making that an emergent property
arises. We accept that emergent properties exists and that they are
the consequence of the interactions of its lower level properties
and the environment --thus we do not accept a compositional physicalism
but an explanatory physicalism \cite{mitchell_EmergAntiKim_2012}
(otherwise obvious in our view). We are rather interested in answering
which are the properties that a complex system may have (and, in particular,
those exhibiting emergent behaviours) for being more or less accessible
to our knowledge applying the scientific method. In this sense, we
align with the proposal of de Haan \cite{de_haan_how_2006}, who highlights
the necessity of a general epistemological framework in which emergence
can be addressed.

An interesting condition for epistemic accessibility was proposed
by Bedau when he coined the term \emph{weak emergence} for those emergent
processes that are epistemologically accessible only by simulation
\cite{bedau_WeakEmergence_1997}. The idea is that a simulation would
demonstrate the \emph{supervenience} of upper level properties from
lower level constituents, even if the mechanistic process leading
to the observed pattern is not completely understood, i.e. it is not
possible to compress the simulation into a compact set of rules explaining
how the outcome is determined. Therefore, it provides an objective
definition of emergence based on computational incompressibility,
that has been explored by different models such as cellular automata
\cite{crutchfield_evolRevol_2003,bedau_downward_2002} or genetic
algorithms \cite{holland_ChaosOrder_2000}, approximations that were
later called \emph{computational emergence} \cite{huneman_DynamEmergence_2008}. 

Nevertheless, Bedau pointed out as well that, even if there is computational
incompressibility, it could be possible to recover from the simulations
regularities that may lead to describe the system under compacts laws
\cite{bedau_WeakEmergence_1997}. Otherwise, the above computational
approximations should be understood as non-epistemic, as they would
be useless to decipher the principles governing the emergence of a
property. Furthermore, Huneman rightly emphasized that this is an
important issue to understand the relationship between computational
models and processes observed in nature \cite{huneman_LessonComputEmerg_2012}
and, for those computational models depicting regularities in their
global behaviour, coined the term \emph{robustly emergent}.

There is a last notion of emergence we would like to discuss that
has been considered fundamental --as opposed to epistemological--,
which is called \emph{strong emergence} \cite{bar-yam_mathematical_2004}.
In Physics it is accepted that knowing the positions and velocities
of particles is sufficient to determine the pairwise interactions.
This assumption is frequently found in Physics-inspired models of
collective behaviour, where individual motion results from averaging
responses to each neighbour considered separately. Nevertheless, Bar-Yam
reasoned that this assertion would not hold if the system is embedded
in responsive media --such as the motions of impurities embedded in
a solid--, or in any process where global optimization (instead of
local) is involved. In this way, if there is in the system a constraint
acting on every component simultaneously and it is strong enough --i.e.
it is a global constraint--, it is not possible to determine the state
of the system considering only pairwise interactions. In some sense,
the parts are determined \emph{downwards} from the state of the whole,
being the consciousness the most paradigmatic example suggested for
strong emergence. 

In this paper, we are interested in understanding which are the conditions
that a natural system depicting an emergent property may have for
fitting the notion of robust emergence. However, we will shift the
attention from computational models to focus on the analysis of experimental
data. We will follow the view in which emergence is considered a ``relation
between descriptions of models of natural systems and not between
properties of an objective reality in itself'' \cite{bich_Synthese_2012}
although, as we will see immediately, we will not renounce to talk
about ontology. In particular, we aim to understand the relation between
macroscopic descriptions of emergent behaviours and their microscopic
explanatory models. Starting from experimentally characterized microscopic
states associated to a macroscopic emergent observation, we want to
understand which kind of regularities found in these microstates are
more difficult to compress and why. We believe that this is a necessary
\emph{a priori} step for any computational approach aiming to model
the observed process, and thus scientifically relevant notions of
emergence should be derived from this process rather than from the
computational models themselves.

An immediate risk in this endeavour is that we must select a framework
to work with and, by doing this, any result may depend on the framework
selected. To circumvent this difficulty, we will apply a framework
founded in logic whose limits are thus clearly established. In particular,
we will analyse with logical principles
the map between the conceptual setting build by an observer following
the scientific method and the objects observed. As a consequence,
we will be able to clearly see why and how this limits are surpassed,
and this map will help us to recover the ontological meaning of these
limitations.

The article is articulated as follows. In Section
2 we introduce the formalism. We carefully provide a definition of
system, introducing next the different mathematical tools needed and
finally providing a topological notion of conceptual vagueness. We
will show here that well known difficulties discussed around the concept
of emergence \cite{emmeche_explaining_1997,ryan_emergence_2007,seth_measuring_2010,de_haan_how_2006,bedau_WeakEmergence_1997,machta_paramCompress_2013},
can be understood through this notion of vagueness. Thus, our effort
in the application of a novel formalism becomes justified because
we find a more expressive picture of the epistemological problems
we face when dealing with complex systems and their ontological origins. We
analyse in Section 3 the macroscopic and microscopic descriptions,
showing how the macroscopic descriptions are more prone to generate
vague concepts while the (bottom-up) microscopic description is used
to build explanatory frameworks. It is in Section 4 where we will
show how the knowledge subject proceeds to build such a framework,
and we discuss how concept disjunction helps us in the identification
of the system's constraints. Then we go into three synthetic systems
in which, applying the new formalism, we identify which are the properties
that make the system more or less epistemologically accessible. We
finish in section 5 discussing the limitations of our approach, and
how to elucidate which is the complexity of the constraints underlying
the system under analysis despite of these limitations, what lead
us to propose a definition of emergence strength.

\section{A topological description of the phase space induced by measurable
properties}

In the following sections we introduce a novel application of topological
notions, whose novelty relies on its ability to formally describe
epistemological questions that are hardly addressed by other approximations.
A nice introduction for computational scientists to the generalization
of the approach presented here, called formal topology, can be found
in \cite{sambin_SomePointsFormalTop_2003}, and a relevant application
to the epistemological determination of what should be understood
as a vague concept is found in \cite{boniolo_vagueness_2008}.

\subsection{System definition}

We start proposing a glossary of terms concerning the system definition,
some of them close to those proposed by Ryan in \cite{ryan_emergence_2007}.
We will call (object of) observation $o_{i}$, to a set of basic magnitudes
associated to a given entity. Each of this magnitudes is a function
$f_{M}$ of the Cartesian product of a collection of $M$ sets --where
at least one of them is determined by an experimental measurement--,
into real numbers $\mathbb{R}$, i.e. $f_{M}:A\times B\times...\times M\rightarrow\mathbb{R}$.

The non measurable sets may refer to a set of measurement units (grams,
meters,...) to a set of reference frameworks, or any other set necessary
to determine the final magnitude. For simplicity, we will consider
that any variation in the magnitudes is a consequence of a variation
in the outcome of a measurement and thus, in the following, we will
not distinguish between magnitude and measurement when objects of
observation are discussed.

In this way, we will consider that our system is characterized by
a bunch of $M$ quantitative and/or qualitative (i.e. binary) magnitudes
$X=\{x_{k},k=1,..,M\}$. Given that we are interested in complex systems,
we will consider that our system consists of a large number of entities,
that we denote with $N$. We will call\emph{ scope }to this selection
of objects whose size, $N$, implicitly determines the spatio-temporal
boundaries of the system. Determining the scope is already a difficult
task for large dynamical systems. These difficulties arise, on the
one hand, from the identification of these entities because, when
the number is large, a complete characterization may be unfeasible.
On the other hand, it will be also difficult to define the separation
between system and environment, as this separation cannot be achieved
many times using strictly objective arguments \cite{alley_OrganismEnvEpistem_1985,maturana_autopoiesisAndCognition_1991}.
We will discuss these questions in more detail below.

The variables selected $X$ are intended to be sufficient to answer
the questions addressed in the research. For simplicity in the exposition,
we start considering an ideal scenario in which all these variables
can be quantified for any entity within the system, leading to $N\times M$
specific values. This assumption will not affect our conclusions,
as we can assign a vanishing value to any variable from which the
associated magnitude is not observed for one or several entities.
We will propose a procedure to relax this assumption in \textcolor{black}{section
}\ref{sub:Coverage-excess} to discuss how the complexity of a system
can be explored following the scientific method.

Every variable $x_{k}$ has a \emph{resolution} $r_{k}=f:x_{k}\rightarrow\mathbb{R}$,
which is the finest interval of variation that we set for that variable,
and it is established from different arguments. For instance,
the resolution may be limited by the intrinsic error in the measurement,
which would be an ontological limitation. Another possibility arises
when the expected influence of a given variable on the system's description
is small for a given shift in the value, and a coarser discretization
is then justified (an epistemological limitation). Calling $I_{k}$
to the domain of the function $f:x_{k}\rightarrow\mathbb{R}$, the
number of possible values considered for the variable $x_{k}$ will
be $\zeta_{k}=I_{k}/r_{k}$. We will call resolution $R$ of the system
to the finest variation that allow us to distinguish two states of
the system, $R=\max_{k}(\{\zeta_{k}\})\ (k=1,..,M)$. 

This choice of variables together with the set of viable values will
be called the \emph{focus} $F$ of the knowledge subject, upper bounded
by $F\sim M\times R$. We finally call the \emph{scale }to the set
of specific values $\{N,M,R\}$. A factor multiplying any of these
values represents a change in the scale of the scope (if we modify
$N)$, or the focus (if we modify $M$ or $R$).\textcolor{black}{{}
Note that, following this definition, the scale is an ontological
attribute as determined by $N$, but it also depends on the epistemological
attributes determined by $M$ and $R$}. Therefore, the breadth of
the focus is very much influenced by epistemological choices\textcolor{black}{.
Interestingly, it has been suggested that emergent behaviours (theories)
are the consequence of a change in the scope \cite{ryan_emergence_2007}
(and not in the focus \cite{machta_paramCompress_2013}).}

\subsection{Measurable properties, concepts and their extension.}

Let us start introducing some definitions, most of them already provided
and justified in \cite{boniolo_vagueness_2008}, that we recover here
for completeness. For the sake of simplicity we will start considering
that our objects of observation $o\in O$ are the components of a
complex system at a given time, i.e. we focus on a single microstate
$\mu$ with $N$ components described by $M$ variables with resolution
$R$. Each of these components is what we consider for the moment
an object of observation. We will move later towards a description
where each object of observation is a microstate, becoming the whole
space of objects the observed phase space. All the definitions considered
in the following for a single microstate can be extended for other
objects with a different scale.

\subparagraph*{Definition: We call a\emph{ basic concept }or\emph{ characteristic}
$c_{a}=x_{k}^{*}$ to the specific value $x_{k}^{*}$ of a variable
$x_{k}$, out of the $\zeta_{k}$ possible values, measured over an
object of observation $o$.}

In this way, if we consider two different measurements of our variables
for the same entity, each of them will constitute a different object
of observation.

\subparagraph*{Definition: We call \emph{focus} $F$ to the whole set of characteristics
considered by the observer: $F=\{x_{k}^{l};\ k=1,..,M;\ l=1,..,\zeta_{k}\}\equiv\{c_{a};\ a=1,..,\tilde{M}\}$,
with $\tilde{M}=M\times\sum_{k}\zeta_{k}$.}

We make explicit here the discrete nature of the conceptual setting
and the relation between resolution and focus, which achieves a suitable
description in terms of characteristics, in turn leading to the definition
of concepts. Note that discreteness here is not an arbitrary choice,
because we are working with experimental values that have associated
an experimental error, making the measurements essentially discrete.
The transition into continuous descriptions is a posterior formal
abstraction made during the modelling process.

\subparagraph*{Definition: We call a \emph{concept} $\nu$ to any non-empty finite
subset of $F$: $\nu=\{c_{1},...,c_{P}\}$, with $P\leq\tilde{M}$.}

We defined the intension of concepts. Given that a concept may contain
a single characteristic, $\nu=\{c\}$, any characteristic can be considered
a concept as well. The distinction between concept and characteristic
will be needed in certain circumstances to indicate a qualitative
difference that makes a concept a more elaborated entity. For instance,
a characteristic may be the outcome of a measurement and a concept
could be the result of that measurement together with its units. Therefore,
the term characteristic will be used to talk about
elementary concepts (such as single measurements outcomes) but, apart
from its utility in making precise the definitions, this distinction
is not necessary for our purposes and, for the sake of a simplified
exposition, both terms will be used interchangeably.

\subsection{Binary operations}

From the previous definitions it is immediate to propose binary operations
to build new concepts.

\subparagraph*{Definition: (Conjunction of concepts). Let $\nu_{1}=\{c_{1},...,c_{P}\}$
and $\nu_{2}=\{d_{1},...,d_{Q}\}$ be two concepts. Then, the conjunction
of $v_{1}$and $v_{2}$ is the concept}

\begin{equation}
\nu_{1}\wedge\nu_{2}=\{c_{1},...,c_{P},d_{1},...,d_{Q}\}\label{eq:conjuction}
\end{equation}

The conjunction of concepts is, in turn, a concept which consists
on the set of all the characteristics contained in both concepts.
Alternatively, we may want to extract, given two concepts, the common
characteristics they share:

\subparagraph*{Definition: (Disjunction of concepts). Let $\nu_{1}=\{c_{1},...,c_{P},b_{1},...,b_{L}\}$
and $\nu_{2}=\{d_{1},...,d_{Q},b_{1},...,b_{L}\}$ be two concepts.
Then the disjunction of $v_{1}$and $v_{2}$ is the concept}

\begin{equation}
\nu_{1}\vee\nu_{2}=\{b_{1},...,b_{L}\}\label{eq:disjunction}
\end{equation}

The disjunction of concepts leads to a concept containing the set
of all characteristics common to both concepts. Note that the set
of concepts we consider determines a partition of the focus and it
will induce, in turn, a partition in the set of objects of observation.
In other words, understanding the relationship between these partitions
requires to determine a constitutive relationship between any single
characteristic belonging to the focus $F$ and the set of objects
$O$. The following constitutive relationship will express that the
objects become cognitively significant by means of the characteristics
measured and, in turn, by the concepts we build from them:

\subparagraph{Definition: (Constitution relation). Let $F$ be the focus over a
set $O$ of objects. Given $o\in O$ and $\nu\in F$, we introduce
a binary relation, $\Vdash$, that we call \emph{constitution relation},
such that by $o\Vdash\nu$ we mean that $\nu$ is one of the concepts
constituting $o$.\\}

With the constitution relation we determine how the objects of observation
are expressed via the conceptual apparatus of the knowing subject.
In addition, we would like to know which objects are constituted by
a given concept:

\subparagraph*{Definition: (Extension of a concept). Let $\nu\in F$ be a concept.
Then the extension $\Ext$ of $\nu$ is the subset of objects of $O$
constituted by $\nu$, that is}

\begin{equation}
\Ext(\nu)=\{o\in O\ |\ o\Vdash\nu\}\label{eq:extension}
\end{equation}

We note here that an immediate consequence of Eq. \ref{eq:extension}
is that any object of observation has necessarily associated a concept,
i.e. it is just cognitively accessible by means of the conceptual
apparatus of the knowing subject. This assertion, if accepted in general,
leads to a Kantian epistemological positioning \cite{boniolo_vagueness_2008}.
In our case, it is a consequence of the fact that our objects of observation
are built from measurements of a reproducible experimental setting,
and hence it is true by construction. Nevertheless, the opposite is
not true as we may deal with concepts for which no object is observed,
i.e. $\Ext(\nu)=\emptyset$. These concepts are considered for instance
if we have \emph{a priori} expectations of the viable values of the
system\footnote{For instance, we know that a group of birds can fly following any
direction in the space even if we systematically observe that they
follow a single direction. From the point of view of the scientific
method, concepts build from \emph{a priori} expectations are very important,
as they may be used to propose null hypothesis which, in general,
can be formulated as $H_{0}:"\nu\ \textrm{observed"}$. It is when
we reject the hypothesis through experiments when we acquire a scientific
knowledge of the process analysed, i.e. that $\nu$ is not observed.}.

Finally, we aim to know what is the extension of a subset $U$ of
concepts $U=\{\{\nu_{1}\},...,\{\nu_{L}\}\}$.

\subparagraph*{Lemma: Let $U$ be a subset of the set $F$ of concepts. Then, the
extension of $U$ is defined by setting}

\begin{equation}
\Ext(U)=\bigcup_{\nu\in U}\Ext(\nu)\label{eq:subset}
\end{equation}

Hence, if we consider two concepts $c_{\text{1}}$ and $c_{2}$, we
should not confuse a concept built by conjunction of concepts $c=c_{1}\wedge c_{2}$
with the subset of concepts $C=\{\{c_{1}\},\{c_{2}\}\}$. In the former
case, we look for objects containing both concepts and, thus, the
number of such objects is smaller or equal than the number of objects
described by $c_{1}$ or $c_{2}$. On the other hand, the subset $C$
extends over objects containing \emph{any} of the concepts, being
its extension the union of the extension of both concepts. In this
paper, we will analyse sets of objects and we will be interested not
only in finding a description that would allow us to ``talk'' about
them but, in addition, we will look for minimal descriptions, i.e.
descriptions containing the lowest number of concepts. We will show
below that disjunction is the basic logical operation we need to obtain such
representations.

\subsection{Topology and vagueness}

We introduce now some more definitions and a theorem, which represent
the basis of the topological approach \cite{boniolo_vagueness_2008}.

\subparagraph*{Theorem 1: If the map $\Ext$ satisfies the extension condition, then
the family $\{\Ext(U)|U\subseteq F\}$ is a topology over the set $O$,
where $U$ is a subset of concepts of the focus $F$.\\}

This map is central in our arguments. The basic characteristics are
defined in terms of measurements over specific objects, and thus the
extension provides a map between these characteristics and the sets
of objects. If we call power set $\wp$ to the set containing all
the possible partitions in which a given set (in our case $O$) can
be divided, a topology will be some subset of the power set containing
a collection of sets called \emph{open sets} --which include the empty
set and the whole set--, and verifying: 1) the arbitrary union of
open sets is another open set in the topology; 2) The binary intersection
of open sets is also another open set in the topology. Therefore,
a topology is a subset of $\wp$ which is \emph{closed} under arbitrary
union and binary intersection of the open sets it contains.

What we are expressing is that, once we have a conceptual setting
built from measurements, the extension function induces a partition
in the set of objects, and this partition fulfils the conditions
for being a topology. In this way ,we can take advantage of the topological
notions of open and closed sets. Justification for the following definitions
can be found in \cite{valentini_binaryPos_Unpublished,boniolo_vagueness_2008}

\subparagraph*{Definition: (Open set) Let $A$ be a subset of the set $O$ of objects.
Then $A$ is an open set if it coincides with its interior $\mathrm{Int}(A)$,
where}

\begin{equation}
\mathrm{Int}(A)=\{o\in O\ |\ (\exists\nu\in F)\ o\Vdash\nu\ \&\ ext(\nu)\subseteq A\}\label{eq:interior}.
\end{equation}

\subparagraph*{Definition: (Closed set) Let $A$ be a subset of the set $O$ of
objects. Then $A$ is a closed set if it coincides with its closure
$\Cl(A)$, where}

\begin{equation}
\Cl(A)=\{o\in O\ |\ (\forall\nu\in F)\ o\Vdash\nu\Rightarrow(\exists o\in O)\ o\in ext(\nu)\ \&\ o\in A\}\label{eq:closed}.
\end{equation}

\subparagraph*{Definition: (Border) Let $A$ be a subset of the set $O$ of objects.
Then the border $\Bd(A)$ of $A$ is the set}

\begin{equation}
\Bd(A)=\Cl(A)\cap\bar{A}\label{eq:border}
\end{equation}

where $\bar{A}$ stands for the complement of the set $A$ with respect
to $O$.

\subparagraph{Definition: (Vagueness) Let $\nu$ be any concept and $U$ be any
set of concepts. Then $\nu$ is a vague concept if $\Bd(\Ext(\nu))$
is non-empty, and $U$ is a vague set of concepts if $\Bd(\Ext(U))$
is non-empty. \\}

As we anticipated, this definition of vagueness will help us to understand
the origin of ambiguities associated to the definition of emergence.

\section{Vagueness and descriptions of the system}

\subsection{Microscopic and macroscopic descriptions}

Following the definitions introduced we aim now to differentiate two
types of variables providing a description of the system at different
scales: microscopic and macroscopic. Note that with microscopic we
do not mean ``atomistic'', we just talk about a significantly shorter
spatio-temporal scale of observation of the system. A particular feature
of the interplay between both scales is that, when a macroscopic property
is observed during the dynamical evolution of a system, even if the
microscopic variables are continuously changing, the macroscopic variables
remain invariant. 
{} 

In the following, we will call microstate $\mu$ to a vector containing,
at a given time, the values of a set of variables $\{x_{k}\}$ that
fully determines the state of the microscopic objects, i.e. $\mu=\{x_{k}^{*}\}$
where $x_{k}^{*}$ stands for a particular value of the variable $x_{k}$.
Therefore, the basic objects of observation we are considering now
are the microstates $o\equiv\mu$. When a coarse graining of the microstates
at the spatial, temporal or both dimensions is performed, it may be
possible to determine macroscopic variables $y_{k}$ describing the
state of the system in the new (coarser) scale. We will call macrostates
to the objects described at the macro scale $o\equiv\xi$. In some
cases, the macroscopic variables $y_{k}$ can be obtained applying
a surjective map $f$ over the microscopic variables $f(x_{k})\rightarrow y_{k}$.
For instance, if we deal with an incomplete (statistical) microscopic
description of an ensemble $P(\mu)$, we can obtain a coarse determination
of a macroscopic variable $y_{k}$ averaging the correspondent microscopic
variable $x_{k}$ \textcolor{black}{weighted by the statistical probability
of the microstates over the ensemble }$\left\langle x_{k}(\mu)P(\mu)\right\rangle $.
But in many other situations it is not possible to find out such mapping,
and we argue here that this fact underlies the problems surrounding
the study of emergent properties: we observe a macroscopic property
such as the collective behaviour of many interacting elements, and
it does not seem possible to explain it from lower levels of description
(for instance from the properties of the entities themselves). 

It is important to underlie that macrostates and microstates definitions
are relative to the scale of observation and they may change if we
move from one scale of description to another. Consider a system described
within a certain temporal scale by a set of microstates $\{\mu_{i}\}$
which are associated to the observation of a single macrostate $\xi$.
Assume now that the system evolves under a sufficiently long path
such that we observe different macrostates and we store $T$ snapshots
of this dynamics, leading to an ensemble of macrostates $\{\xi_{u}\}_{u=1}^{T}$.
It is possible to consider that each of these macrostates is now a
microstate $\hat{\mu}$ for a new system with a larger scope and lower
resolution $\xi_{u}\rightarrow\hat{\mu}_{i}$. Given that the scope
of a macrostate will be always larger than that of a microstate ($N_{\xi}\geq N_{\mu}$),
whereas it occurs the opposite with the resolution ($R_{\xi}\leq R_{\mu}$),
in this exercise we have increased the scope and reduced the resolution.
This is the reason why, larger is the scale, more difficult
is to build a bottom-up explanatory framework. This movement along
different scales will be very relevant when evolutionary systems are
considered, given that we will need to distinguish at least two spatial
and temporal scales. For instance, such a change is needed when moving
from the ecological analysis of few individuals to the evolutionary
analysis of entire populations. 

Note as well that this change in the scale requires an effort to reduce
the system description, but this kind of reduction has been performed
from the very first step: for the definition of scope, we have neglected
entities; for the focus, we have neglected variables and probably
restricted their viable values assuming a lower resolution. Furthermore,
any map between microstates and macrostates again considers a reduction
in the information provided by the microstates. In general, for both
very broad or very detailed questions the technical complexity increases
and a reduction in the description is unavoidable, and it is important
to remark that this exercise does not mean that the approach is reductionist.
Reductionism should be considered an epistemological attitude where
it is accepted that any macroscopic description is a simple extrapolation
of the properties of the microscopic description \cite{anderson_MoreIsDiff_1972}.
Instead, we accept that in complex systems there are discontinuities
between the different levels of description and that, for each new
level, new properties may arise. We are interested here in investigating
which are the minimum conditions to say that a microscopic description
is a valid representation of an emergent macroscopic observation.

\subsection{Vagueness in the macroscopic description and dialectic concepts \label{sub:Macroscopic-and-dialectic}}

In this section, we aim to justify with a simple example a plausible
origin for the controversies around emergent properties. Importantly,
we will not provide any\emph{ a priori} definition of emergent property.
For the moment, we just assume that there are processes that are sufficiently
surprising for any observer as to qualify them as emergent. We argue
that, given that the macroscopic scale is less refined in terms of
number of variables and values, it is more prone to generate dialectic
concepts. Therefore, we expect that it is in the macroscopic scale
where emergent properties are first identified. %

Let's start considering a simple example where we investigate a complex
system for which two different concepts can be built macroscopically.
The first concept $\hat{c}_{\Omega}$ describes the performance of
the complex system when it is able to visit every possible viable
values of the focus, and thus the subscript $\Omega$ indicates that
it describes a behaviour found for any microstate of the phase space,
i.e. it is not constrained. Remember that we use the \emph{hat} over
the concept to denote that it is a macroscopic concept. For concreteness,
let us assume that the concept is $\hat{c}_{\Omega}=\textrm{"groups of birds flying"}$.
Now consider that, from time to time, we observe that these birds
depict a swarming (or flocking) behaviour and we coin an specific
concept for this $\hat{c}\E=\textrm{"flocking"}$, where the subscript
$E$ stands for ``emergent'', just because we find the behaviour
novel and appealing. If we call $\{o\E\}$ to the observations where
we appreciate a flocking behaviour and $\{o\NE\}$ to those where
we do not, the conceptual setting describing the system will be:

\[
\begin{array}{cc}
\Ext(\hat{c}_{0})= & \emptyset\\
\Ext(\hat{c}\E)= & \{o\E\}\\
\Ext(\hat{c}_{\Omega})= & \{o\E,o\NE\}.
\end{array}
\]

where $\hat{c}_{0}$ is the concept describing the null observation
of the system. According with the previous definitions, $\Bd(\hat{c}\E)=\{o\NE\}$
and we identify $\hat{c}\E$ as a vague concept. The fact that the
border of the concept is the set of observations \emph{not} depicting
flocking $\{o\NE\}$ behaviour means that we are still not good
at explaining why these observations do not belong to the set of observations
that do show flocking behaviour $\{o\E\}$. As a consequence, the
set $\{o\NE\}$ cannot be \emph{safely separated} from $\{o\E\}$.
The reason is that the concept $\hat{c}\E$ is still not informative
of what ``flocking'' means and, as a consequence, we cannot use
it to say what does not mean, what would help us in separating both
sets (i.e. we cannot properly build a concept $\hat{c}\NE=\neg\hat{c}\E$,
where $\text{\textlnot}$ is the NOR operator). This is indeed an
important question to work under the scientific method, because we
need to establish hypothesis of the type $H_{0}:\textrm{"}d\textrm{ observed"}$
which, upon rejection, lead to the concept $c=\text{\textlnot d}$
($d$ is not observed)\textcolor{red}{.}

David Bohm pointed out that, \textcolor{black}{in the earlier stages
of any science, the interest is focused on ``the basic qualities
and properties that define the mode of being of the things treated
in that science'' \cite{bohm_causality_1971}, being tasks such as
comparative analysis and classification the cornerstones in its earlier
development. We argue, that it is in a descriptive (exploratory) context
in which the larger number of vague concepts arise. }Following the
classification of concepts proposed by Georgescu-Roegen \cite{georgescu-roegen_entropy_1971}
we will refer to concepts containing any source of vagueness as dialectic.
To this type belong breadth concepts \cite{stock_concepts-rels_2010},
those related with dynamical properties like stability \cite{ives_stabilityRev_2007},
the difference between organism and machine \cite{bunge_emergence_2003}
or the exact intension of function, autopoiesis and complexity \cite{emmeche_explaining_1997}.
These are classical examples of dialectic concepts and it is remarkable
the potential these concepts have for generating debate, which should
be considered an asset \cite{strunz_resilience-vagueness_2012}.

\textcolor{black}{To continue with Bohm's view on scientific evolution,
it is just after a sufficient exploitation of the dialectic knowledge
when we will find a growing interest on ``processes in which things
have become what they are, starting out from what they once were,
and in which they continue to change and to become something else
in the future'' \cite{bohm_causality_1971}}. \textcolor{black}{This
knowledge, following again the classification of concepts proposed
by Georgescu-Roegen, is built on arithmomorphic concepts: \textquotedblleft {[}arithmomorphic
concepts{]} conserve a differentiate individuality identical in all
aspects to that of a natural number within the sequence of natural
numbers''}\cite{georgescu-roegen_entropy_1971}\textcolor{black}{.
Arithmomorphic concepts are suitable for formal reasoning and quantitative
treatment}, and we argue that emergent processes are dialectic macroscopically,
and that intense research is developed around them upon a successful
microscopic arithmomorphic scheme is built.

\subsection{Microscopic description and arithmomorphic scheme}

Microscopically, when $\{o\NE\}$ is observed the system depicts
a dynamical evolution where constraints are absent. Constraints here
should be simply understood as limitations in the viable values of
the variables we handle \cite{hooker_Constraints_2013}. Any system
is constrained in some extent. But there are some constraints that
belong to the definition of the system itself, that we will call \emph{intrinsic},
and others that depend on particular conditions to be observed, that
we will call \emph{facultative}. Think in the structural differences
between a protein, and a heteropolymer whose sequence is the result
of randomly shuffling the protein sequence. Although both chains of
amino-acids depict the same number of intrinsic constraints (those
derived from the existence of peptidic bonds), a protein structure
requires three additional constraints levels: the first is needed
for being kinetically foldable in a biologically relevant time, the
second for making the fold thermodynamically stable under physiological
conditions, and the third for performing its specific function (metal-binding,
phosphorilation, etc.). Both chains have the same amino-acids but
the evolutionary process has selected for an specific order in the
sequence that generates the constraints needed to make possible that
the emergent property (the protein function) arises. Quantitatively, the
probability that these constraints appear by chance is quite low:
the number of possible heteropolymers of length $N$, considering
an alphabet of 20 amino-acids, is $20^{N}$ and the number of protein
structures (of\emph{ any} length) deposited at date in the Protein
Data Bank is $\sim1.2\times10^{5}$ (www.pdb.org). If we call $\{\mu\E\}$
the set of microstates where the emergent behaviour is observed, $\{\mu\NE\}$
the region where it is not observed, and $\Omega$ the whole phase
space, we expect that $\#\{\Omega\}\approx\#\{\mu\NE\}\gg\#\{\mu\E\}$;
where $\#\{\text{·\}}$ is the cardinality of the set, i.e. the number
of elements it contains. Therefore, we expect that the volume of the
region of the phase space were an emergent property arises to be much
smaller than the whole phase space. This is probably why unpredictability
or surprise are attributes frequently used for emergent properties.

Recovering the example of Sec. \ref{sub:Macroscopic-and-dialectic}
we know that the following map between the macroscopic description
and the underlying processes exists:

\[
\begin{array}{cc}
\Ext(\hat{c}_{0})= & \emptyset\\
\Ext(\hat{c}\E)= & \{\mu\E\}\\
\Ext(\hat{c}_{\Omega})= & \{\{\mu\NE\},\{\mu\E\}\}.
\end{array}
\]

Again, the concept $\hat{c}\E$ determines an open whose closure
is the whole phase space $\Cl(\Ext(\hat{c}\E)=\{\mu\E\})=\{\{\mu\NE\},\{\mu\E\}\}$.
The intersection between the closure and this open determines its
border $\Bd(\Ext(\hat{c}\E)=\{\mu\E\})=\{\mu\NE\}$ which
is non-empty which, as we have seen, means that the concept $\hat{c}\E$
is a vague concept. Getting into the microscopic description, vagueness
will vanish if we build a topology such as:

\[
\begin{array}{cc}
\Ext(c_{0})= & \emptyset\\
\Ext(c\E)= & \{\mu\E\}\\
\Ext(c\NE)= & \{\mu\NE\}\\
\Ext(c_{\Omega})= & \{\{\mu\NE\},\{\mu\E\}\}
\end{array}
\]

where $c\NE=\text{\textlnot}c\E$. In the examples we develop
in the next sections, we will call generically to the positive (null)
observation with the concepts $c$ ($d$). For all the systems analyzed,
we will assume that some microstates correspond to a macroscopic emergent
observation, and vagueness removal will be achieved when we are
able to find a subset of concepts $U$ such that $\Ext(U\NE)=\{\mu\NE\}$,
which means that, when only the emergent process is observed $\Ext(U\NE)=\emptyset$,
we reject the hypothesis $H_{0}:"U\NE\text{ is observed}"$.
Therefore, in the following section we aim to investigate how we can
build a microscopic arithmomorphic description looking for microscopic
concepts that allow us to differentiate between the sets of microstates
$\{\mu\E\}$ and $\{\mu\NE\}$.

\section{A synthetic approximation to emergent properties}

\subsection{Traceability and compact descriptions}

With the above considerations, we propose two formal definitions that
will be helpful to understand our rational behind the further development
of the paper. We first define what is considered a novel macroscopic
property, which identification is typically the starting point of
any research.

\subparagraph{Definition: (Novel macroscopic property). We will say that an observed
macroscopic property is a novel property if it is observed only in
the presence of certain facultative constraints limiting the viable
values of the system.\\}

Therefore, given that the phase space of the system $\Omega$ is restricted
to a smaller observed region $\Omega\Ob\subset\Omega$, what we say
is that there exists a macroscopic concept $\hat{c}$ such that $\Ext(\hat{c})=\Omega\Ob$,
and we would like to explore this region both in terms of macroscopic
$\{\hat{c}\}$ and microscopic $\{c\}$ concepts. We now introduce
a condition that allow us to consider that a macroscopic description
is in correspondence with a microscopic description.

\subparagraph*{Definition: (Traceability). Given a novel macroscopic property $\hat{c}$
and the observed phase space $\Omega\Ob$ associated to that property,
i.e. $\Ext(\hat{c})=\Omega\Ob$, we will say that the macroscopic
description obtained is traceable if we find an appropriate function
or algorithm applied on microscopic properties $f:\{c\}\rightarrow q$
such that the new concept $q$ derived \emph{compactly} describes
the ensemble of microstates, i.e. $\Ext(q)=\Ext(\hat{c})=\Omega\Ob$.\\}

This definition paves the way for quantifying the correspondence between
both descriptions within the framework proposed.
Note that it does not require to be able to relate macroscopic and
microscopic properties, but only to establish a correspondence between
microscopic and macroscopic variables describing the same region of
the observed phase space $\Omega\Ob$. Therefore, traceability can
be seen as a rather minimal epistemological condition, because it
is what allow us to talk about emergent properties circumventing any
epistemological discontinuity between both descriptions. Still, we
need to clarify what is understood by compact description.

\subparagraph*{Definition: (Compact description) We say that a set of distinguishable
microstates $\{\mu\}$, namely a set in which every microstate $\mu$
is completely described through the microscopic set of concepts $F$, is compactly
described by a concept $q$, if $\Ext(q)=\{\mu\}$ and the number of
concepts needed to build $q$ is strictly smaller than $F$.}

In the next sections we show with different examples how can be achieved
a compact description in order to say that a novel macroscopic property
is traceable.

\subsection{Identification of constraints: focusing on disjunction}

In this section, we would like to investigate if there is any general
method to reach a compact description of the ensemble of microstates.
As we said, when an emergent property is observed, the system is constrained
to a certain region of the phase space. This means that there is a
breaking of symmetry,\textcolor{black}{{} namely the probability distribution
for values of the different variables depart from the distribution
observed when the system is free of constraints, thus losing ergodicity
\cite{auyang_CsystTheories_1999} (p. 186). Therefore, the existence
of facultative external or internal constraints limit the behaviour
of the system and, as we will attempt to clarify, a necessary condition
for determining a microscopic property associated to every microstate
visited requires the determination of the existing constraints. We
will see that the nature of the different constraints acting on the
system determine its epistemological accessibility and, thus, our
ability to reach a satisfactory explanation of emergent behaviours.}

Firstly, we show how can be obtained the extension of those concepts
built through binary operations over sets of concepts. When we obtain
a new concept $\tau$ via conjunction, for instance $\tau=\nu_{1}\wedge\nu_{2}$,
the extension of the new concept will be the intersection of the sets
of objects associated to each of the starting concepts $\Ext(\tau)=\Ext(\nu_{1})\cap \Ext(\nu_{2})$.
Aiming to fully identify a single object requires to determine a sufficiently
large number of concepts in order to sharply separate it from the
other objects, being conjunction the basic operation that permits
to reach more precise descriptions. 

\begin{figure}
\includegraphics[width=1\textwidth]{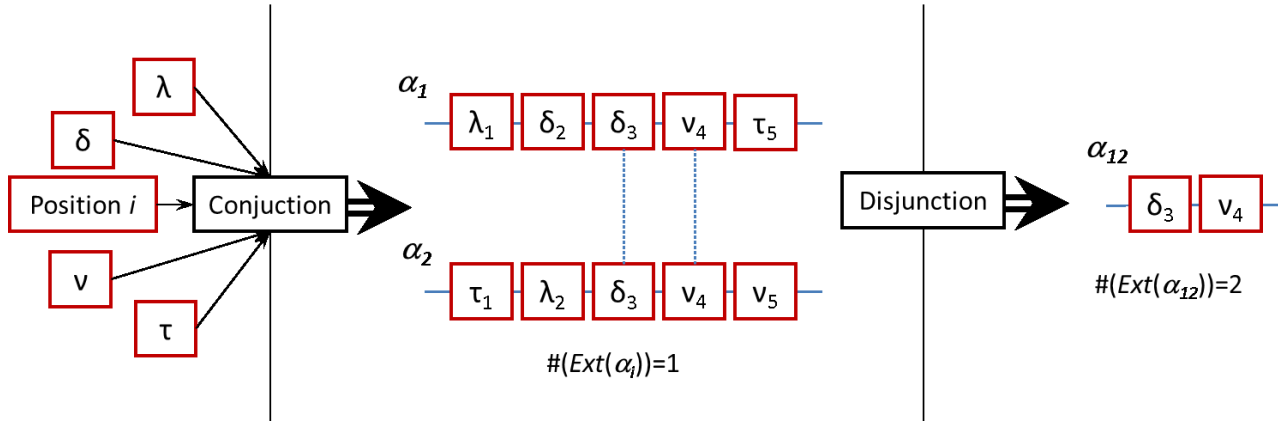}

\caption{\label{fig:ConjDisj} \textbf{Illustration of conjunction
and disjunction of concepts. }Starting from the knowledge subject's
conceptual apparatus (greek letters, left), two sequences $\alpha_{1}$ and $\alpha_{2}$
are built through conjunction of the basic concepts, being themselves
concepts (center). These sequences uniquely determine a single object,
for instance a protein sequence, and thus $\#(\Ext(\alpha_{i}))=1$.
By comparison of both sequences we observe two common concepts (linked
by dotted lines) that we extract through binary disjunction leading
to a concept $\alpha_{12}$ (right) containing less basic concepts
but whose extension is larger than the original sequences ($\#(\Ext(\alpha_{12}))=2$)
being its scope larger. In the case of proteins it may be understood
as a signature of their common ancestry, i.e. of their homology.}
\end{figure}

Let us take as an example the description of a set of proteins $\{o_{\alpha}\}$
provided by the sequence of their amino-acid composition, which is
embedded within an evolutionary phase space. Each amino-acid molecule
in the protein is a component of the system which is described by
its position in the sequence and by a single variable whose specific
value consists of one out of the $20$ natural amino-acids encoded
by DNA. In this way, an example of concept within this description
would be something like \emph{$\nu_{i}=$``cysteine in position $i$''
--}which, in turn, is built by conjunction of the more basic characteristics
``\emph{cysteine}'' and ``$i$''\emph{--}. A protein sequence
$o_{\alpha}$ will be subsequently built by conjunction of a set of
such a kind of concepts describing the amino-acid observed at each
position, i.e $\alpha=(\nu_{1}\wedge\nu_{2}\wedge...\wedge\nu_{N})$
(see Fig. \ref{fig:ConjDisj}). The sequence becomes uniquely determined
under this description, i.e. the extension of the sequence maps exactly
one object of observation, namely, the protein under study: $\Ext(\nu_{1}\wedge\nu_{2}\wedge...\wedge\nu_{N})=\Ext(\alpha)=o_{\alpha}$.
In summary, conjunction underlies bottom-up approximations, where
we focus in an accurate description through the compilation of concepts.

Let us now have a look to another kind of concepts $\lambda$, which
are obtained via disjunction, $\lambda=\nu_{1}\vee\nu_{2}$. Following
the equation \ref{eq:disjunction}, given that $\nu_{1}=\{c_{1},...,c_{P},b_{1},...,b_{L}\}$
and $\nu_{2}=\{d_{1},...,d_{Q},b_{1},...,b_{L}\}$, the extension
of the concept $\lambda=\{b_{1},...,b_{L}\}$ will be given by $\Ext(\lambda)=\Ext_{i<j}(b_{i}\wedge b_{j})$
(see Fig. \ref{fig:ConjDisj}).\textcolor{black}{{} From this definition
we note that $\Ext(\{\{\nu_{1}\},\{\nu_{2}\}\})=\Ext(\nu_{1})\cup \Ext(\nu_{2})\subseteq \Ext(\nu_{1}\vee\nu_{2})$.
Hence, this is an operation that allow us to find commonalities, which
may be extended to objects that are not included in the sets of objects
over which the concepts $\nu_{1}$ and $\nu_{2}$ are extended.}\textcolor{green}{{}
}Disjunction stands out as a relevant operation to look for breadth
concepts, and it is consistent with the intuition stating that these
concepts tend to overtake the boundaries of our starting focus.

\subsection{The three bits system}

We are already equipped with all the tools necessary for analysing
in detail a synthetic example. The toy model we consider consists
on a system of three entities whose physical state is described by
a single binary variable, i.e. a system modelled with three bits.
From an experimental point of view, there may be three distinguishable
entities (ranging from a molecule to a population) described with
binary variables. We can think in sets of genes that are expressed
(not expressed) when the amount of the correspondent protein is above
(below) certain threshold, species observed (absent) in certain environmental
sample or, from a strictly computational experiment, the attractor
of a boolean network. Each measurement performed over these entities
will be considered an observation, and each of them may take a value
of one or zero. For a system composed by three entities we can observe
$2^{3}=8$ microstates $\mu_{k}=(x_{1},x_{2},x_{3})$ (with $k=1,..,8;$
and $x_{i}=\{0,1\}$; see table \ref{tab:microstates}). As the three
entities can be distinguished the focus is
\begin{lyxlist}{00.00.0000}
\item [{$c_{1}='ON\ at\ object\ 1';\ d_{1}='OFF\ at\ object\ 1';$}]~
\item [{$c_{2}='ON\ at\ object\ 2';\ d_{2}='OFF\ at\ object\ 2';$}]~
\item [{$c_{3}='ON\ at\ object\ 3';\ d_{3}='OFF\ at\ object\ 3';$}]~
\end{lyxlist}
Each microstate is defined in terms of this focus through concepts
$e_{k}\ (k=1,...,8)$ built by conjunction of characteristics. For
instance, the microstate $\mu_{7}=(1,1,0)$ is defined in terms of
the basic characteristics as $e_{7}=c_{1}\wedge c_{2}\wedge d_{3}$,
being in turn a concept. The fact that we work with elementary concepts
build from measurements in which both the observation or not of certain
property can be assessed experimentally, stands on the basis of the
construction of a bottom-up explanatory framework. The observation
of a ``flocking behaviour'' cannot be assessed experimentally
unless we clearly determine measurable values. A bottom-up characterization
in terms, for instance, of the relative angles in the positions between
birds allow us to tackle the vagueness associated to the term
``flocking''. Furthermore, it has been claimed that flocking is
not observed in a single microstate \cite{ryan_emergence_2007}. However,
once we have characterized flocking with microscopic variables, for
example once we derive the pairwise distribution of interactions among
neighbours \cite{bialek_flocking_2012}, we can also test whether
any behaviour, even for a single snapshot in the flight of a group
of birds, is consistent with the model derived. This fact already
provides the means to build an hypothesis testing experiment where
the null hypothesis is that the group of birds do not present flocking
behaviour. 

\begin{table}
\caption{\label{tab:microstates} Microstates of a three-bits system.}

\centering{}%
\begin{tabular}{|c|c|}
\hline 
\multicolumn{2}{|c|}{Microstate}\tabularnewline
\hline 
\hline 
\textbf{$\mu_{1}=(1,1,1)$} & \textbf{$\mu_{5}=(0,1,1)$}\tabularnewline
\hline 
\textbf{$\mu_{2}=(1,0,0)$} & \textbf{$\mu_{6}=(1,0,1)$}\tabularnewline
\hline 
\textbf{$\mu_{3}=(0,1,0)$} & \textbf{$\mu_{7}=(1,1,0)$}\tabularnewline
\hline 
\textbf{$\mu_{4}=(0,0,1)$} & \textbf{$\mu_{8}=(0,0,0)$}\tabularnewline
\hline 
\end{tabular}
\end{table}

We can define for the three-bits system $\ensuremath{\dbinom{N}{n_{\mathrm{ext}}}}$
possible combinations of constraints involving $n_{\mathrm{ext}}$ variables,
and thus the number of final microstates will depend on the number
of constraints and their scope, i.e. the number of elements influenced
by the constraint. In the following, we consider examples
with a different number and type of constraints, all resulting in
the same number of microstates (four out of the eight viable states).
Hence, these constraints are codified in one bit of information, but
we will see that the number of concepts needed to express these constraints
can change from system to system. 

The most simple macroscopic description associated to the observed
ensemble, arises if we consider a coarse graining of the microscopic
properties such that there is a surjective map between microstates
and macrostates a macroscopic variable takes the value $'ON'$ if these microstates
are visited and $'OFF'$ otherwise. In this way, only if there is
a statistically significant bias towards these microstates we can
say that a novel emergent macroscopic property is observed.

Taking these considerations in mind, we aim to disentangle the microscopic
constraints in the system following our formalism. Given that we build
our conceptual setting starting from the basic characteristics (obtained
from measurements) and then performing binary logical operations,
we expect that the results obtained for the different systems are
fairly comparable. We will perform this comparison taking into account
the number of propositions found and the concepts contained, i.e.
analysing whether the respective representations provide a more or
less compact description. Therefore, a valid description for the constraints
should represent a reduction of dimensionality of the system (a compact
description), as it is a necessary condition to build any simplified
model.

\paragraph*{System with a single constraint of scope one (S1).}

The rational is the same for the following three systems. We consider
that there is an observed macroscopic emergent observation $\hat{c}\E$,
and we know the microstates associated to that particular region of
the phase space $\{\mu\E\}$. Then, we analyze the set of microstates
looking for its constraints.

The first system we consider is a system where the first bit is constrained
to a fixed value ($c_{1}$), leading to the observations $\{\mu_{1},\mu_{2},\mu_{6},\mu_{7}\}$
that we explicitly show in table \ref{tab:Single}.

\begin{table}
\caption{\label{tab:Single} Three bits microstates associated to the region
of the phase space where an emergent property was observed $\{\mu\E\}$,
for a system with a single constraint of scope one.}

\centering{}%
\begin{tabular}{|c|}
\hline 
\multicolumn{1}{|c|}{$\{\mu\E\}$}\tabularnewline
\hline 
\hline 
\textbf{$\mu_{1}=(1,1,1)$}\tabularnewline
\hline 
\textbf{$\mu_{2}=(1,0,0)$}\tabularnewline
\hline 
\textbf{$\mu_{6}=(1,0,1)$}\tabularnewline
\hline 
\textbf{$\mu_{7}=(1,1,0)$}\tabularnewline
\hline 
\end{tabular}
\end{table}

\begin{figure}
\includegraphics[width=1\textwidth]{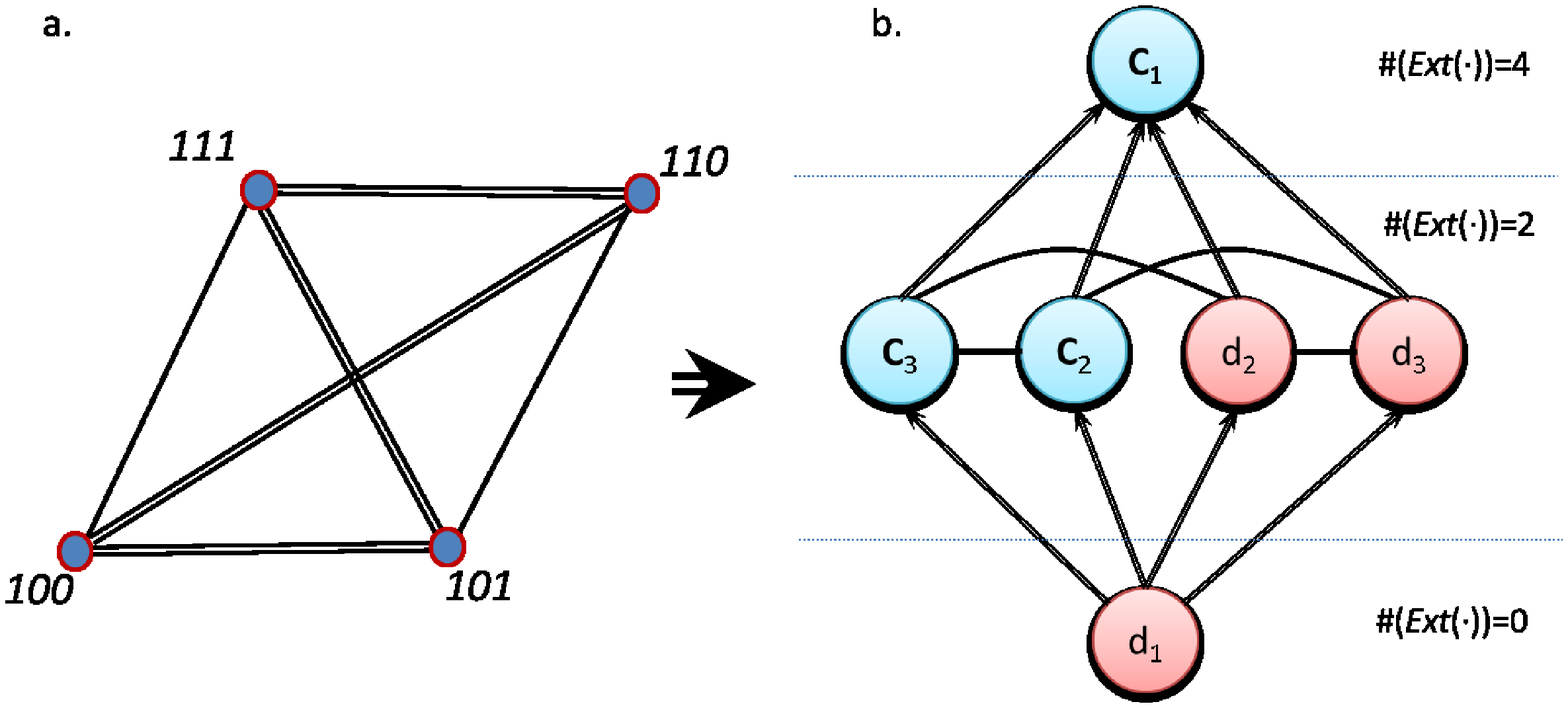}

\caption{\label{fig:OneConstraintExtOne}\textbf{Representations
of a three bits system with a single constraint of scope one.} (Left)
In the network, each node represents a microstate and it is linked
with another microstate if they share the same observation for any
component, where the number of links represent the number of concepts
shared. (Right) Network of concepts extracted from the analysis of
the microstates. Two links $c_{i}$ and $c_{j}$ are linked with a
directed edge if $\Ext(c_{i})\subseteq \Ext(c_{j})$ and with an undirected
link if $\Ext(c_{i})\cap \Ext(c_{j})\protect\neq\emptyset$. The concepts
are hierarchically ordered according to the cardinality of their extension,
i.e. number of microstates they map. In this example a single constraint
on $x_{1}$ naturally arises, as one of its possible values maps the
empty set.}
\end{figure}

In order to find the system constraints we start comparing the concepts
$e_{i}$, which determine the different microstates. We provide a
compact representation of these comparisons with a network, see Fig.
\ref{fig:OneConstraintExtOne}, where each concept $e_{i}$ is linked
with a concept $e_{j}$ if they share a basic concept, $c$ or $d$.
Although the constraints determine the microstates, these act on the
variables so we need to go one step further to identify them. We move
from a network of microstates to a network of basic concepts, and
we link two concepts $c_{i}$ or $d_{i}$ if they extend onto the
same microstates (see Fig. \ref{fig:OneConstraintExtOne}). More formally,
we link two concepts $c_{i}$ and $c_{j}$ with a directed edge if
$\Ext(c_{i})\subseteq \Ext(c_{j})$, and with an undirected edge if
$\Ext(c_{i})\cap \Ext(c_{j})\neq\emptyset.$ In this way, we compactly
represent all the dependencies present in the system represented with
relationships of subordination (directed edges) cooccurrence (undirected)
or exclusion (absent link) between the different values. This representation
resembles the information that we would recover if we build a network
of variables from a covariance matrix: positive (negative) correlation
arises when similar (dissimilar) values are found between two objects

From the network relating variables in Fig. \ref{fig:OneConstraintExtOne}
it is easy to observe that one of the values of the first variable,
$d_{1}$, is never observed, a fact that we can express with the proposition:

\[
\Ext(d_{1})=\emptyset
\]

What this proposition simply states is that, in order to identify
that a given microstate belongs to this system, it is necessary to
evaluate that the value measured on the first component of the system
is different from zero. In other words, as soon as we reject the hypothesis
$H_{0}:\textrm{"}d_{1}\text{ is observed"}$, we will be confident
in that we are dealing with a microstate contained in $S1$. Note
that we achieved a quite important reduction of the number of concepts
needed to talk about the system: from the four concepts $\{e_{1},e_{2},e_{6},e_{7}\}$,
each of them containing three basic concepts, we reduce to a subset
containing a single concept $U\E=\{c_{1}\}$. In addition, the set
of non-emergent microstates are well described by the subset $U\NE=\{d_{1}\}$,
what prevents the description for being vague.

\paragraph*{System with two constraints of scope two (S2).}

\begin{figure}
\includegraphics[width=1\textwidth]{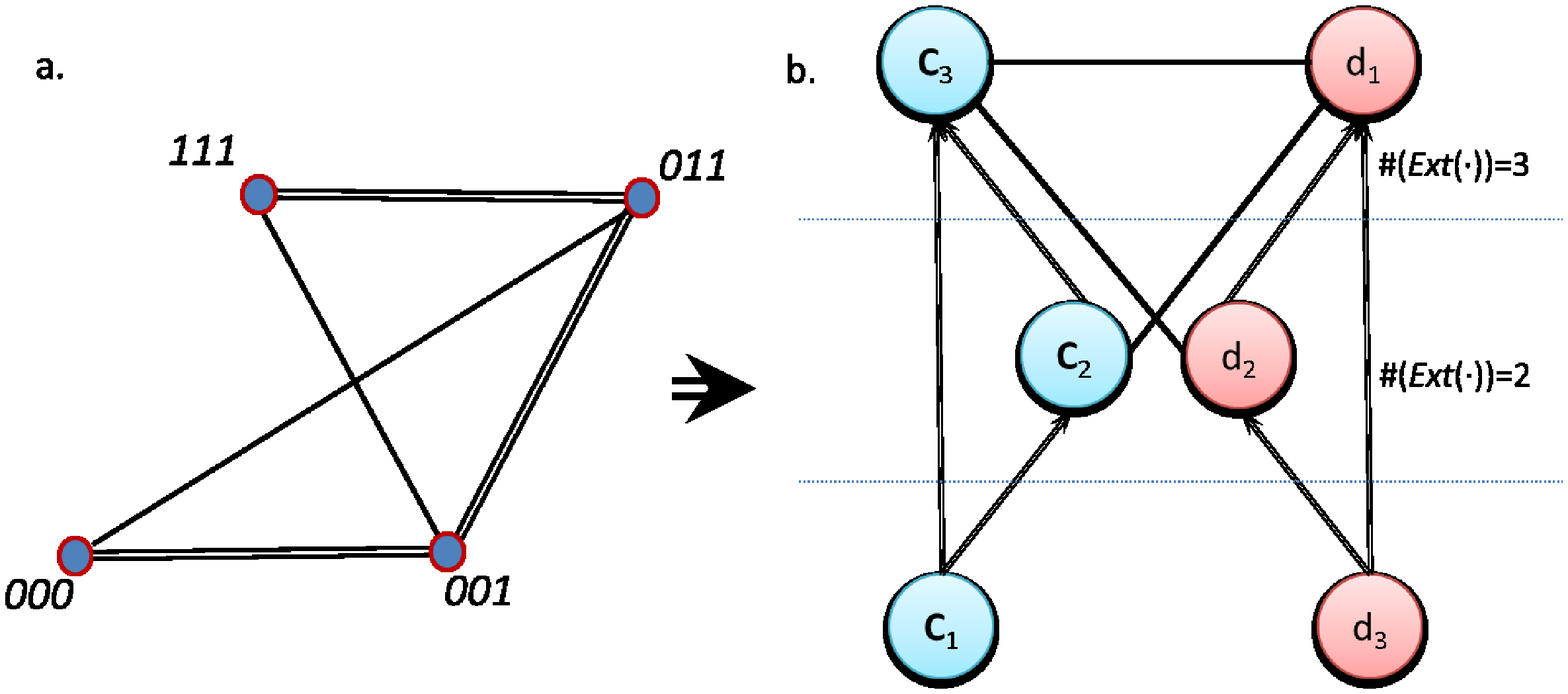}

\caption{\label{fig:ThreeConstraintsExtTwo}\textbf{ Representations
of a three bits system with two constraints of scope two.} (Left)
In the network, each node represents a microstate and it is linked
with another microstate if they share the same observation for any
component, where the number of links represent the number of concepts
shared. (Right) Network of concepts extracted from the analysis of
the microstates. Two links $c_{i}$ and $c_{j}$ are linked with a
directed edge if $\Ext(c_{i})\subseteq \Ext(c_{j})$ and with an undirected
link if $\Ext(c_{i})\cap \Ext(c_{j})\protect\neq\emptyset$. The concepts
are hierarchically ordered according to the cardinality of their extension,
i.e. number of microstates they map. In this example we identify the
constraints observing those links that being viable are absent, for
instance there is no link between $d_{3}$ and $c_{2}$.}
\end{figure}

We select now four microstates that are obtained imposing one constraint
among each pair of variables. Taking the microstates $\{\mu_{1},\mu_{4},\mu_{5},\mu_{8}\}$
(that are explicitly shown in table \ref{tab:Linear}), and repeating
the procedure of the previous example (see Fig. \ref{fig:ThreeConstraintsExtTwo}),
we observe that the disconnected components in the graph lead to the
following constraints, which can be expressed with the propositions:

\[
\begin{array}{cc}
\Ext(c_{1}\wedge d_{2})= & \emptyset\\
\Ext(c_{\text{2}}\wedge d_{3})= & \emptyset\\
\Ext(c_{1}\wedge d_{3})= & \emptyset
\end{array}
\]

\begin{table}
\caption{\label{tab:Linear} Three bits microstates associated to the region
of the phase space where an emergent property was observed $\{\mu\E\}$,
for a system with two constraints of scope two.}

\centering{}%
\begin{tabular}{|c|}
\hline 
\multicolumn{1}{|c|}{$\{\mu\E\}$}\tabularnewline
\hline 
\hline 
\textbf{$\mu_{1}=(1,1,1)$}\tabularnewline
\hline 
\textbf{$\mu_{4}=(0,0,1)$}\tabularnewline
\hline 
\textbf{$\mu_{5}=(0,1,1)$}\tabularnewline
\hline 
\textbf{$\mu_{8}=(0,0,0)$}\tabularnewline
\hline 
\end{tabular}
\end{table}

It is easy to observe that one of these constraints is redundant.
Given that $c_{2}$ and $d_{2}$ cannot be observed simultaneously,
if $c_{1}$ is observed it means that $c_{2}$ is also observed and
thus $d_{3}$ cannot be observed. And the other way around, if $d_{3}$
is observed $c_{2}$ will not be observed and thus $c_{1}$ cannot
be observed. Therefore, the third constraint $\Ext(c_{1}\wedge d_{3})=\emptyset$,
is a consequence of the other two. We formulate our result positively
saying that the set of emergent microstates $\{\mu\E\}$ is described
by the subset $U\E=\{\{d_{1}\wedge c_{2}\},\{d_{2}\wedge c_{3}\}\}$,
and the set $\{\mu\NE\}$ by $U\NE=\{\{c_{1}\wedge d_{2}\},\{c_{2}\wedge d_{3}\}\}$

\paragraph*{System with a single constraint of scope three (S3; the parity bit
system).}

Our last example is a set of microstates having an even number of
$ON$ bits, i.e. a single constraint involving all three components.
This system has been previously introduced by Bar-Yam as a toy example
of the particular type of emergent behaviour we introduced above called
\emph{strong emergence} \cite{bar-yam_mathematical_2004}. For this
system, given that we find two random values in two randomly selected
bits, the third bit is constrained in such a way that the number of
bits in the microstate is always odd. This rule is used in the control
of message transmission, where the last bit (called the parity bit)
is used to monitor the presence of errors in the chain transmitted.
Note that we are not interested in understanding the system under
this engineering perspective, as it provides already a rather \emph{ad
hoc} explanation on how the system is built \cite{bersini_Synthese_2012}
and, in this work, we assume no \emph{a priori} knowledge of the underlying
mechanisms generating the observation. It is just one possible observation
that will be analysed as in the previous examples. The microstates
we will consider are $\{\mu_{1},\mu_{2},\mu_{3},\mu_{4}\}$, explicitly
shown in table \ref{tab:Parity}

\begin{table}
\caption{\label{tab:Parity} Three bits microstates associated to the region
of the phase space where an emergent property was observed $\{\mu\E\}$,
for a system with a single constraint of scope three.}

\centering{}%
\begin{tabular}{|c|}
\hline 
\multicolumn{1}{|c|}{$\{\mu\E\}$}\tabularnewline
\hline 
\hline 
\textbf{$\mu_{1}=(1,1,1)$}\tabularnewline
\hline 
\textbf{$\mu_{2}=(1,0,0)$}\tabularnewline
\hline 
\textbf{$\mu_{3}=(0,1,0)$}\tabularnewline
\hline 
\textbf{$\mu_{4}=(0,0,1)$}\tabularnewline
\hline 
\end{tabular}
\end{table}

\begin{figure}
\includegraphics[width=1\textwidth]{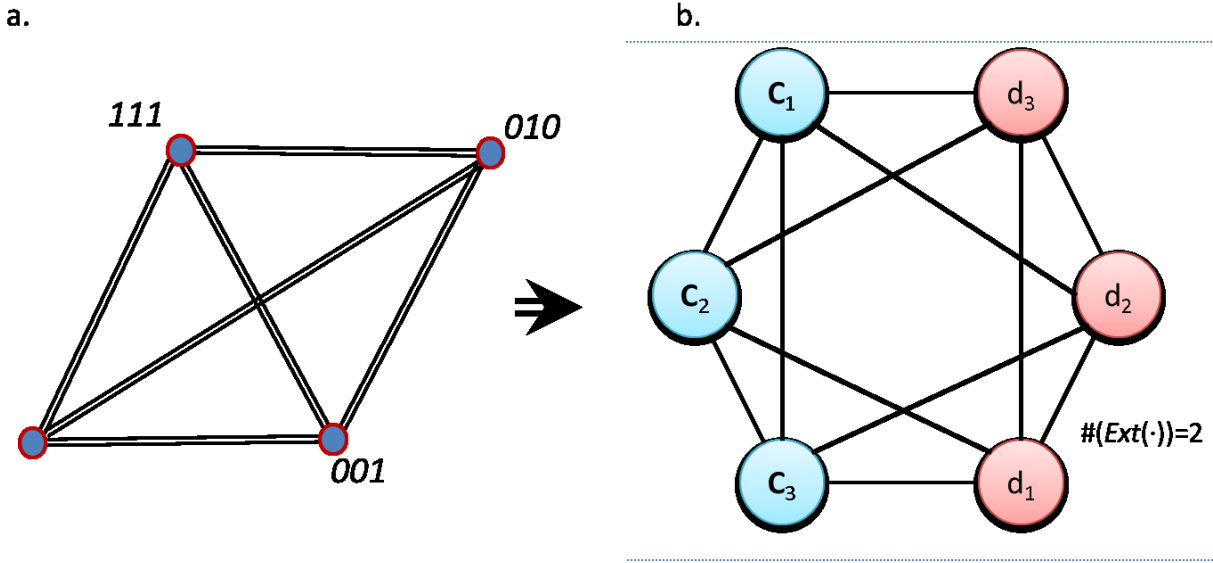}

\caption{\label{fig:OneConstraintExtThree}\textbf{Representations
of a three bits system with one constraint of scope three.} (Left) In
the network, each node represents a microstate and it is linked with
another microstate if they share the same observation for any component,
where the number of links represent the number of concepts shared.
(Right) Network of concepts extracted from the analysis of the microstates.
Two links $c_{i}$ and $c_{j}$ are linked with a directed edge if
$\Ext(c_{i})\subseteq \Ext(c_{j})$ and with an undirected link if $\Ext(c_{i})\cap \Ext(c_{j})\protect\neq\emptyset$.
The graph of concepts is equivalent to the graph we would obtain for
a free system, being just observed a reduction in the number of objects
mapped by each concept (from $\#\Ext(\text{·)=4}$ towards $\#\Ext(\text{·)=2}$).
It reflects the notion that the system has ``no borders'', and no
constraints can be explicitly extracted with our framework.}

\end{figure}

In this case, the network of concepts intuitively resembles an sphere
in the sense that there are no ``borders'' --i.e. disconnected concepts
from which propositions about the constraints are simply derived (see
figure \ref{fig:OneConstraintExtThree})--. Thus, the identification
of constraints is possible only because we already know the viable
values. Indeed, a parallel analysis of the free system highlights
a lower cooccurrence of the different variable values, but there will
be no differences in the final network topology we obtain. This fact
would be also observed in the set of marginal probability distributions,
as no bias will be observed for the free system nor for the parity
bit system.

The comparison with the free system bring to the surface the following
propositions:

\[
\begin{array}{cc}
\Ext(d_{1}\wedge c_{2}\wedge c_{3})= & \emptyset\\
\Ext(c_{1}\wedge d_{2}\wedge c_{3})= & \emptyset\\
\Ext(c_{\text{1}}\wedge c_{2}\wedge d_{3})= & \emptyset\\
\Ext(d_{\text{1}}\wedge d_{2}\wedge d_{3})= & \emptyset
\end{array}
\]

And what we observe is that the most compact way to talk about this
system within this formalism is to write down all the microstates
that are \emph{not} observed. Therefore, we obtain no reduction of
dimensionality at all what will make difficult to build any model.
In this way, vagueness in the macroscopic concept $\hat{c}\E$ still
holds, because microstates from $\{\mu\NE\}$ are getting attracted
by those $\{\mu\E\}$. More technically, $\Bd(\Ext(\hat{c}\E)=\{\mu\E\})=\{\mu\NE\}$,
again because we do not consider a valid solution for solving vagueness
of $\hat{c}\E$ simply describing it through the subset $U\E=\{e_{1},e_{2},e_{3},e_{4}\}$,
because it is not a \emph{compact description}.

\section{Emergence}

\subsection{The three-bits system in other representations }

In the previous examples, we showed how our formalism worked differently
depending on the type of constraints present in the system. We found
that it apparently fails in getting a description of the constraints
if there is a single constraint with a scope equal to the system's
size. Of course, we may think that our formalism is simply insufficient
and that there may be other more sophisticated formalisms that would
be able to express the constraints in S3, with a much lower amount
of information that the one we need to describe it. For instance,
there may be a function $q$ such that, given a description of a microstate
$e_{i}$, returns a new \emph{type} of concepts $\tilde{c}=q(e_{i})$
that are able to differentiate the microstates of S3 from the rest.
Let's see this with an example. We can characterize the three examples
above with their probability distributions:

\begin{eqnarray*}
 & P(\mu)=\delta(x_{1},1)/2^{n+1} & (S1)\\
 & P(\mu)=\delta(\delta(H_{12}+1,1),H_{23}+1)/2^{n+1} & (S2)\\
 & P(\mu)=\delta(\mathrm{mod}_{2}(\sum_{i}x_{i}),1)/2^{n+1} & (S3)
\end{eqnarray*}

where $x_{i}$ is the value of the bit $i$,\textcolor{black}{{} $n$
is the number of bits, $\delta(\text{a,b)}$ is the Kronecker's delta,
$H_{ij}=H(x_{i}-x_{j}-1)$ is the Heaviside function and $mod_{2}(\text{·})$
is the module two function. In principle, there is no reason to think
that the probability distribution of S3 is more complex than those
of S1 and S2. For saying that it should be any objective reason to
say, for instance, that the use of a summation operator and a module
function is more complex than applying two Heaviside functions. }

\textcolor{black}{A fairer comparison can be assessed through the
algorithmic information complexity (AIC) \cite{gellmann_InfoMeasures_1996},
also known as Kolmogorov complexity. We approximate this measure programming
three simple scripts (one for every system) that generate strings
of size $N$ containing these constraints. In the Appendix, we provide
the pseudocode for these programs along with some technical details,
and the scripts can be found in the Supplementary Material. Compiling
this scripts lead to three binaries which, after compression, have
the following ratios of compression 3.070:1Kb (S1) > 3.011:1Kb (S2)
< 2.998:1Kb (S3). With these results, the constraint implemented in
S3 requires more information to be codified than those in S2 and S1,
suggesting that it is a more complex constraint. Nevertheless, there
is no dramatic difference as to identify it as particularly relevant
or complex. In addition, in the Appendix we discuss that there are
some choices in the way we implemented this exercise that may lead
to different results, such as the ability of the subject writing the
code.}

What we aim to point out with these examples is that, the fact that
the formalism we are using here has established limits, is not a drawback
but an advantage to clearly set the boundaries of the framework. For
instance, a desirable property of a framework should be that, if we
increase the system's size, our ability to describe the constraints
of a larger system should change according with their number and scope.
If we think in how these systems will increase the number of microstates
when the number of components $N$ increases, both S1 and S3 will
increase as $2^{N-1}$ while S2 increases as $N+1$. If we use AIC
to quantify the constraints, the same script can generate a small
or large string just changing the value of the variable $N$. According
with our formalism, the number of propositions needed to describe
the constraints in the system remains equal to one for S1, increases
as $N-1$ for S2, and as $2^{N-1}$ for S3. We believe that this finding
is remarkable and, in the following sections, we show which are the
consequences of this observation, and how can be used to provide a
quantification of emergence compatible with the scientific method.

\subsection{Coverage excess\label{sub:Coverage-excess}}

It is difficult to investigate the effects that an increase on system's
size may have on traceability, because it would mean that we already
have a full knowledge of the underlying constraints and how they would
extend through the new elements, which is not typically the case when
investigating complex systems. An alternative would be to \emph{intervene}
in the system neglecting components of the model and then monitor
which is the relative change in our ability to predict the system's
behaviour after the intervention. This strategy has been highlighted
as a basic ingredient to link computational modelling with the scientific
method \cite{boschetti_intervention_2011}. When we neglect any component,
we will reduce our predictive power if we lose constraints and, thus,
there will be more states of the system compatible with the remaining
constraints. We can quantify which is the uncertainty we generate
when we lose constraints, what means that we can relate the causal
effects that a component has on the other components, according to
the notion of Granger causality \cite{seth_measuring_2010}. 

\begin{figure}
\centering{}\textcolor{black}{\footnotesize{}\includegraphics[scale=0.6]{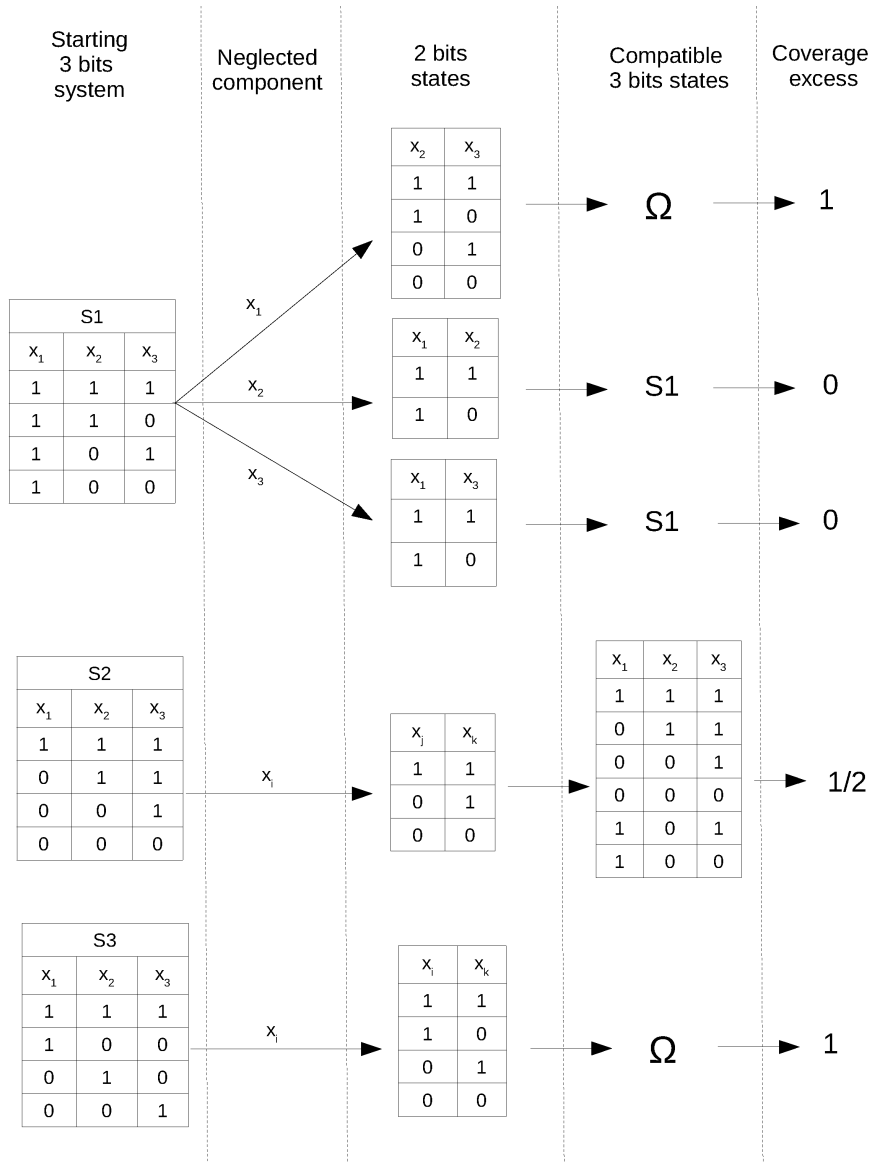}\caption{\textcolor{black}{\emph{\label{fig:CoverageExcessExample} }}\textcolor{black}{Scheme
illustrating the definition of coverage excess. The scheme is divided
in five columns (1-5) that we describe from left to right. (1) The
three-bits systems under analysis in the main text are shown. If we
intervene in the systems neglecting one component (2) we will obtain
a set of 2-bits states (3). For the system S1, removing $x_{1}$ lead
to different states than if $x_{2}$ or $x_{3}$ are removed, while
S2 and S3 lead to the same states independently of the component removed
(see Main Text for details). From the two-bits states, we recover
the neglected component keeping it free of any constraint, which leads
to a number of compatible 3-bits states (4). In the last column we
show the result for the coverage excess obtained from this procedure
using Eq. \ref{eq:Coverage Excess}. The final value for the coverage
excess of the system will be the average among the values obtained
from the different interventions.}}
}
\end{figure}

We illustrate the proposal in Fig. \ref{fig:CoverageExcessExample}
considering the three-bits synthetic examples again. Starting from
any of the systems, we neglect the components one by one, and we explore
which are the two-bit states recovered. Depending on whether any constraint
was lost or not, we are able to build from these states a number of
three-bit states. We simply introduce back into the system the variable
neglected allowing it to get any viable value. Exploring systematically
all the variables we infer which is the influence of the underlying
constraints. For instance, for S1, if we neglect the first component
$x_{1}$ --where the constraint relies--, we obtain all possible states
of a two-bits system containing components $x_{2}$ and $x_{3}$.
This is because the unique constraint in the system was deleted, and
thus we can recover the whole phase space $\Omega$ when we build
up all the three-bit states compatible with these two-bit states.
On the other hand, if we neglect any of the other two components $x_{2}$
or $x_{3}$, the constraint still remains in the first component $x_{1}$.
Therefore, if we look into the three-bit states compatible with the
two-bit states we are constrained for coming back to the original
system S1. For the system S2, irrespective of the component neglected
we reach the same two bits states. From them, we obtain not only the
original three bits states but two more, what means that we have lost
one constraint every time, but the other one is still active. For
S3, irrespective of the component neglected, we obtain all possible
two bits states, and hence the whole phase space will be always recovered,
hence the constraint is always completely lost. 

With this kind of interventions we can quantify how the original system
is \emph{covered in excess }when any intervention takes place. More
formally, let's call $U$ to the subset of concepts contained in the
focus $F$ describing a set of microstates $S=\{\mu\}$ containing
$N$ components in the phase space $\Omega$, which we know is associated
to a novel macroscopic property. Then remove any of the $x_{i}$ components
and consider a new set of microstates $S'$ of a system with $N-1$
components. Next call $V(x_{i})$ to the subset of concepts describing
the set of microstates $S''$ of a system with $N$ components obtained
by adding a new unconstrained component $x_{i}'$ to $S'$. We say
that the coverage excess $\Xi$ of $S$ induced after neglecting the
variable $x_{i}$ and introducing the component $x'_{i}$ is the quantity

\subparagraph*{
\begin{equation}
\Xi(x_{i})=\frac{\#(\Ext(V(x_{i})))-\#(\Ext(U))}{\#(\Ext(F))-\#(\Ext(U))}=\frac{\#(S'')-\#(S)}{\#(\Omega)-\#(S)}\label{eq:Coverage Excess}
\end{equation}
}

where the function $\#(\text{·})$ returns the number of microstates
contained in the set. This quantity takes a value of zero, when the
states recovered are the same than the original ones, and it takes
one when it is recovered the whole phase space. To consider a single
value for the coverage excess we will average the result of the intervention
over all variables:

\begin{equation}
\left\langle \Xi\right\rangle =\frac{1}{N}\sum_{i}\Xi(x_{i})\label{eq:AvCoverageExcess}
\end{equation}

If the system is very large an exhaustive computation may be unfeasible,
and a random sampling of the different components 
or more complex interventions such as the removal of several variables,
and we indicate this averaging over interventions with the brackets
$\left\langle \text{·}\right\rangle $. For the examples explored
in the three bits system, we obtain $\left\langle \Xi\right\rangle _{S1}=1/3$,
$\left\langle \Xi\right\rangle _{S2}=1/2$ and $\left\langle \Xi\right\rangle _{S3}=1$.
The coverage excess reflects the vulnerability of the system to this
intervention and, as the effects have to do with the number of variables
over which we intervene and with the scope of the constraints present
in the system, it provides a mechanism to differentiate between
upward and downward causation \cite{bitbol2012downward}. The system
S1 is very vulnerable if $x_{1}$ is neglected but it is not affected
at all if any other variable is neglected, and thus there is upward
causation from the first variable to the whole system. On the other
hand, the system S3 is very vulnerable, as the coverage excess is
maximum irrespective of the variable over which we intervene, thus
highlighting that there is a global constraint affecting the system
downwards . Note as well that these values will scale differently
with system size. While for $S1$ $\left\langle \Xi\right\rangle _{S1}\rightarrow0$
when $N\rightarrow\infty$, for $S3$ it will remain constant and
equal to one. Finally, we consider the particular case in which we
are already dealing with a subset of microscopic concepts $U$ such
that an emergent process described by the macroscopic concept $\hat{c}$
is perfectly covered, i.e. such that $\Ext(U)=\Ext(\hat{c})$. In this
case, we will say that the Eq. \ref{eq:AvCoverageExcess} provides
\emph{the coverage excess of the emergent property }$\hat{c}$, that
we denote as $\left\langle \hat{\Xi}\right\rangle $.

\subsection{Loss of traceability and emergence strength}

The sensitivity of the knowledge of the observer when it intervenes
on the system suggests that, in the research process, systems with
higher coverage excess will be more difficult to analyse. This difficulty
can be combined with the notion of traceability we proposed above.
If we have a perfectly traceable system, we can quantify how much
we lose traceability after intervention with the \emph{loss of traceability}

\begin{equation}
\Upsilon=1-\left\langle \hat{\Xi}\right\rangle .\label{eq:LossOfTraceability}
\end{equation}

With this quantity we express that those systems which are easily
covered in excess have a low traceability, and the other way around.
Therefore, the associated macroscopic property will be difficult to
explain, from which the following definition for the \emph{emergence
strength} $\sigma$ of the emergent property arises naturally

\begin{equation}
\sigma=-\mathrm{log}(\Upsilon).\label{eq:Emergence strength}
\end{equation}

We took the natural logarithm of the traceability to represent the
emergence strength as a distance between an state of perfect knowledge
of the system (complete traceability) and the state after intervention.
If, after intervention, we still remain in a situation of perfect
traceability, this distance will be zero. On the other hand, if we
completely lose all the information about the system in a way that,
with the above procedure, we recover an unconstrained phase space,
this distance will be infinity, thus reflecting some epistemological
gap for these systems. 

With these definitions we expect to reconcile different positions
on whether the origin of emergence is epistemological or ontological:
even if we deal with a perfectly traceable system --thus epistemologically
accessible--, we can still see that there are systems that are particularly
more inaccessible than others, and there are ontological reasons for
that, namely the type of constraints involved in the system. For these systems,
until perfect traceability is attained, we will probably be tempted
to say that they are epistemological inaccessible, and that there
is an strictly ontological and not epistemological reason for that.
But it is a combination of both, there is an ontological reason why
their emergence strength is so high as to build an epistemologically
accessible (microscopic) description, but it does not mean that such
description cannot be achieved at some point. We propose to solve
controversies around the definition of weak and strong emergent properties
\cite{bar-yam_mathematical_2004} using the emergent strength: if
the emergence strength is infinite we deal with a strongly emergent
pattern whereas if the value is finite we deal with a weak emergent
pattern with the associated strength as an indicator of how difficult
it is to trace it. 

Of course, we cannot discard that there exist systems epistemologically
inaccessible. This may be the case for quantum, relativistic or computational
systems --some of which the definitions of weak and strong emergence
where originally thought--, but not for many systems of scientific
interest (such as living systems) where we believe that the situation
is rather the one that we expose here: these systems are very large
and they are under constraints with a large scope. Note that it may
be argued that, with the above definitions, it cannot be determined
the emergence strength unless we achieved perfect traceability. This
is only true for strong emergent properties because Eq. \ref{eq:LossOfTraceability}
can be easily modified to consider an intermediate state of knowledge,
just using the expression in Eq. \ref{eq:AvCoverageExcess} (and not
$\left\langle \hat{\Xi}\right\rangle $), which will give us an estimation
of the emergence strength. We expect indeed that for natural systems
there is a complex structure of constraints with different scopes,
and we will be able to progressively discover this structure --possibly
from low to high scopes--, and provide an estimation at any time.

\section{Discussion}

In this article we proposed a novel approach to investigate the concept
of emergence in complex systems. We tackled the problem through the
formalism of concrete topology, which is a constructive logical system
that permits the investigation of the relationship between concepts
and objects of observation \cite{boniolo_objects_2012}. In doing
so, we focused in a particular kind of systems, which we believe attain
much interest in the nowadays discussion of emergence. First, we considered
that we are analysing a naturally occurring macroscopic emergent property,
and not a process generated from a computation. In addition, we neglect
any vitalism, what means nothing but accepting explanatory physicalism
--using the words of Mitchell, what else could be?\cite{mitchell_EmergAntiKim_2012}--.
To continue with, we considered that we are able to describe microstates
of the system through experimental measurements. This implicitly assumes
that we are able to differentiate the system from its background \cite{maturana_autopoiesisAndCognition_1991}
and that we are able to provide a bottom-up characterization in terms
of concepts associated to the elements that constitute the system
\cite{bich_Synthese_2012}. Nevertheless, we considered as well that
we have no clue on which mechanistic processes underlie these observations,
as it happens when research on a new process is starting. When the
underlying mechanism is well understood, there are already interesting
proposals in the literature providing measures of emergence \cite{hoel-tononi_MacroBeatMicro_2013},
but obviously this is not the case for most of the natural processes
under research. Finally, we highlighted that there exist a relationship
between the macroscopic observation of the emergent property and the
constrained walk of the system in certain region of the phase space.
With these conditions in mind, we aimed to talk about systems from which we
expect to find sufficient regularities in the analysis of microstates
as to be able to build explanatory models, i.e. in potentially robust
emergent systems, which are the focus of the scientific interest \cite{huneman_LessonComputEmerg_2012}.

Equipped with these tools and with a minimal description of complex
system we showed that, when two descriptions coexist, the less detailed
description is prone to generate dialectical concepts, that are vague
unless a more explanatory description removes the ambiguity. This
is the case between the macroscopic and microscopic description when
an emergent property is observed, as it is typically not immediate to
explain the emergent phenomena from the microscopic description. We showed that, 
building a microscopic model aimed
to remove the ambiguity of the macroscopic observation, requires
to identify constraints in the viable values of the microscopic variables.
Interestingly, we identified concept disjunction as the basic logic
operation to find constraints. Since long it has been recognized the
importance of comparisons for the proper determination of any object,
that may be viewed as a negative determination through the exploration
of the limits of the object, as it was stated by Hegel \cite{hegel_georg_2010}: 
\begin{quotation}
``the object, like any determinate being in general, has the determinateness
of its totality outside it in other objects, and these in turn have
theirs outside them, and so on to infinity. The return--into--self
of this progression to infinity must indeed likewise be assumed and
represented as a totality, a world; but that world is nothing but
the universality that is confined within itself by indeterminate individuality,
that is, a universe.'' 
\end{quotation}
With disjunction we explore the progression of an elementary object
into other objects that may eventually lead to the identification
of new objects exceeding the individuality of the starting objects,
and then back: the identification of objects with a larger scope reinforce
the individuality of the elementary objects. The search of similarity
measures, dissimilarity measures or distances is an essential task
in Biology and Ecology \cite{legendre_numerical_2012} aiming to understand,
following a top-down approach, the information shared between the
different observations. This is probably why methods comparing objects
of observation, such as protein sequence alignments like BLAST \cite{altschul_BLAST_1990},
are the most cited ever in the scientific literature \cite{van_to100CitedPapers}.
In general, disjunction is on the basis of dimensionality reduction
techniques such as principal components analysis \cite{hotelling_PCA_1933}.
Following our framework, these are techniques aiming to obtain a representation
with the minimum number of concepts explaining the maximum variability
in the space of objects. In this way, we are able to \emph{talk} about
the set of objects using a subset of concepts, which is essentially
the task addressed by dimensionality reduction techniques, and that
we defined here with the notion of compact description. 

We then applied these tools to three different ensembles of microstates
of a three-bits synthetic system. We observed that the scope of the
constraints is the main difficulty to identify them: larger is the
scope of the constraint more difficult is to assess it. In particular,
our method was unable to find a compact representation when the scope
of the constraint has the same size than the system, which directly
links the epistemological limitation of our framework with an ontological
property of the system. We briefly visited other approximations, identifying
that the type of constraints heavily influence the consequences that
either an increase in system size or a loss of components may have
in our ability to identify them. This observation seems to be independent
of the formalism followed, and thus of any subjectivity induced by
the one we chose here. 

Notably, we were able to express this observation with concrete topology.
We proposed a procedure based on the intervention of the observer
on the system, thus compatible with the scientific method, to compute
the loss of information experienced when we neglect components in
the system. As this loss of information depends on the type of constraints
present, we can quantify how difficult is to achieve traceability
between the microscopic and macroscopic description. The loss of traceability
was then used as a quantity to establish a distance between perfect
traceability and our knowledge of the system after intervention, what
we called the emergence strength.

We believe that, for the kind of systems we are analysing, the emergence
strength paves the way to reconcile the differences between the notion
of weak and strong emergence. In the systems we are interested, we
aim to develop computational models to reproduce the experimentally
measured data and simulate the emergent process, and thus it is compatible
with weak emergence. Nevertheless, we propose to move the focus from
the ability of building a computational model to simulate the process
to an earlier scientific stage, namely the identification of constraints
from experimental data. In the identification of constraints is where
we start learning about the nature of the natural process we face.
In this way, we focus in understanding the number and scope of the
constraints, whose complexity will determine its emergence strength. 

We conjecture that for systems with different types of constraints,
those with smaller scope are identified first. Accordingly, if a system
has only constraints with a large scope or there is a big gap with
respect to those with lower scope, it may be simply impossible at
a certain stage of knowledge to assess them, an example of this may
be our nowadays knowledge of consciousness. For these processes, the
emergence strength may be so large that it would be justified to call
them strongly emergent processes. This definition seems compatible
as well with the classification proposed by de Haan \cite{de_haan_how_2006},
as the existence of a microscopic emergent conjugated causally
affecting the macroscopic pattern (in the strongest case, consciously), 
can be understood in terms of a global constraint (as he recognises
relating this type of emergence with downward causation). This is the case in living systems
where we believe strong emergence may be pervasive. 

Natural selection can be viewed as a global constraint acting in very
large temporal scales for every individual organisms. The constraints
that one individual feels will have certain similarities with those
felt by another individual (e.g. periodic exposition to day-light)
being others specific to the individual's micro-environment. The term
closure has been coined to described the fact that organisms and the
environment (which includes other organisms) are entangled through
interactions exerting a mutual influence such that natural selection
is modulated by the activity of the organisms themselves \cite{chandler_closure_2000}.
This picture, in its stronger version (in which the influence is so
high that the notion of individual as object of selection is challenged),
becomes more and more important in nowadays research, overall in 
microbial world (see for instance \cite{cordero_MicrobEvol_2014}).

Consider the following example proposed in \cite{pascual_Thesis_2015},
in which we consider one individual for which its fitness $f_{i}$
can be decomposed in two components, where the first component reflects
the fitness $f_{ij}^{int}$ of the individual as a consequence of
its ecological interactions with other species $j$, and the second
its fitness $f_{i}^{\overline{int}}$ due to any other process, thus
$f_{i}=f_{ij}^{int}+f_{i}^{\overline{int}}$. Now consider a particular
example, in which two individuals belong to two different species,
$a$ and $b$, interacting mutualistically. Finally, think in an evolutionary
event which becomes fixed in the population of species $a$ affecting
its fitness, $f_{a}\rightarrow\hat{f}_{a}$, in such a way that the
new fitness $\hat{f}_{a}<f_{a}$ and, in particular, $\hat{f}_{a}^{\overline{int}}=f_{a}^{\overline{int}}$
but $\hat{f}_{ab}^{int}<f_{ab}^{int}$. This means that the fitness
of species $b$ due to the interaction with species $a$ will be also
affected after the evolutionary event and, thus, there will be a change in the selection pressure
of the regions of the genomes of both species codifying the traits
needed for the interaction. Furthermore, if we consider an extreme scenario
in which $f_{ab}^{int}\gg f_{a}^{\overline{int}}$ and $f_{ba}^{int}\gg f_{b}^{\overline{int}}$
--that may be the case for auxotrophs (see a synthetic ecological
experiment in\textcolor{black}{{} \cite{mee-wang_PNAS_2014}})--, the
importance of these coevolving regions in the evolutionary process
would be so high, that the conception of object of selection should
be revisited \cite{mayr_ObjectSelection_1997}: it might be more
appropriate to frame the evolution of both species considering them
as a particular kind of multicellular species. This is probably why
it has been emphasized the importance of mutualistic interactions
in emergent processes \cite{corning_reemergence_2012}, although recent
theoretical results suggests that mutualism is not necessary to derive
a mathematically precise definition of community-level fitness \cite{Tikhonov_elife_2016}.

Not surprisingly, there is increasing interest in the development
of methods for the inference of interactions, but the essence of the
question remains the same. If we aim to understand emergent processes
through interaction patterns, we will compare microstates representated
through networks of interactions, and the main evolutionary constraints will be
identified finding their common topological properties  \cite{pascual_Thesis_2015}.

In summary, we find that the formalism we used here improves our ability
to synthetically understand complex problems. We believe it could
be used as well to face other challenging questions such as the concept
of closure, and thus we hope that our effort will stimulate both the
scientific and philosophical community. Looking for
fresh formal approaches to talk about philosophical questions is particularly
important because, just as formal frameworks help us 
in making predictions, they will help us as well in shaping the philosophical
knowledge and to establish new links between
science and philosophy. This would be probably good news for science,
as the benefits of philosophy seem to be, for nowadays scientists,
left behind.

\section*{Acknowledgements}

I am particularly in debt with Silvio Valentini, who patiently solved
many doubts in the starting stages of this work, and for his kind
hospitality in a short visit to the Universitá degli studi di Padova
(Italy). Part of this work was developed in a short stay in the Centro
Superior de Salud Pública in Valencia (Spain), in the laboratory of
Prof. Andrés Moya, who provided me interesting comments and readings.
Jorge Ibáñez-Gijón brought to my attention several interesting discussions
in the literature and encouraged me to start this project. I acknowledge
to Ugo Bastolla and to several teachers and students of the Santa
Fe Institute Summer School held in El Zapallar (Chile) for fruitful
discussions, in particular to Fernando Rosas for long discussions
about the parity-bit system. This work was supported by grant BFU2012-40020
of the Spanish Government and by an ERC starting grant (project 311399)
awarded to Thomas Bell, to whom I acknowledge his support. Research
at the CBMSO is facilitated by the Fundación Ramón Areces.

\section*{Appendix}

In this appendix we provide the pseudocode to generate $M$ strings
of size $N$ containing the constraints described in systems S1, S2
and S3 in the Main Text. An implementation in Fortran is available
in Supplementary Material. The scripts have been developed trying
to keep similar most of the structure for the three systems, which
can be summarized as follows:
\begin{lyxcode}
\%Variables~declaration~

type~x(N)~\%~describes~every~bit~in~the~system~x=\{0,1\}

for~i~in~1~to~M
\begin{lyxcode}
for~j~in~1~to~N
\begin{lyxcode}
if~constraint
\begin{lyxcode}
x(j)=applyConstraint()
\end{lyxcode}
else
\begin{lyxcode}
x(j)=rndGenerator()
\end{lyxcode}
endif

write~x(j)
\end{lyxcode}
endfor
\end{lyxcode}
endfor
\end{lyxcode}
Where the function \texttt{applyConstraint}\textsf{ }applies the constraint
correspondent to the condition\textsf{ }\texttt{constraint}\textsf{,
}which depends on the system, and the function\textsf{ }\texttt{rndGenerator}\textsf{
}randomly generates a zero or a one. For each system, the constraints
are applied as follows:
\begin{lyxcode}
\%~S1~constraint:

if(j==1)
\begin{lyxcode}
x(j)=1
\end{lyxcode}
endif

\_\_\_\_\_\_\_\_\_\_\_\_\_\_\_\_\_\_\_\_\_\_\_\_\_

\%~S2~constraint:

if((x(j-2)==1)or(x(j-1)==1))
\begin{lyxcode}
x(j)=1
\end{lyxcode}
endif

\_\_\_\_\_\_\_\_\_\_\_\_\_\_\_\_\_\_\_\_\_\_\_\_\_

\% S3 constraint:

if(j==N)

~~~~Remainder=MOD(Total,2)
\begin{lyxcode}
if(Remainder==0)THEN
\begin{lyxcode}
x(j)=1
\end{lyxcode}
else
\begin{lyxcode}
x(j)=0
\end{lyxcode}
endif
\end{lyxcode}
endif
\end{lyxcode}
where \texttt{Total} is a variable that sums-up all the $N-1$ previous
values and\texttt{ MOD} is a function that returns the remainder of
a module two division of \texttt{Total}. Note that, while S1 and S3
have a straightforward generalization of their constraints from a
3-bits system to a N-bits system, it is not the case for S2 where
several generalizations are possible. We decided to extend the system
adding bits from left to right and keeping the original constraints
along the chain. This will generate a bias in the distribution of
the output microstates, rapidly decreasing the probability of generating
strings containing many zeros for increasing $N$.

Next, the scripts have been compiled with gfortran (gcc versión 4.8.5
in Red Hat 4.8.5-4) and the binaries compressed with both bzip2 (version
1.0.6) and gzip (version 1.5), obtaining qualitatively similar results.
It is important to remark that the scripts should be coded with as
few dissimilarities as possible. For instance,
S2 and S3 require two additional variables respect to S1, which already
make a difference in the compression ratios. But, if for whatever
reason, we decide to use for these two variables integer types for
S2 and real types for S3 (or the other way around) the results may
be qualitatively different. This is one of the reasons why the complexity
of the system is difficult to assess with this kind of approximation.
Although our implementation is close to be
optimal, as the size of the binaries before compression is almost
the same (13467Kb for S1 and S2 and 13519Kb for S3)
we find that concrete topology provides a fairer framework for comparing
different systems.

\bibliographystyle{ieeetr}
\bibliography{Pascual-Garcia_Epistemology_arXiv_2016}

\begin{thebibliography}{10}

\bibitem{nietzsche_Zaratustra_2014}
F.~Nietzsche, {\em Thus Spoke Zarathustra}.
\newblock GoodBook LLC, 2014.

\bibitem{mitchell_EmergAntiKim_2012}
S.~D. Mitchell, ``Emergence: Logical, functional and dynamical,'' {\em
  Synthese}, vol.~185, no.~2, pp.~171--186, 2012.

\bibitem{corning_reemergence_2012}
P.~A. Corning, ``The re-emergence of emergence, and the causal role of synergy
  in emergent evolution,'' {\em Synthese}, vol.~185, no.~2, pp.~295--317, 2012.

\bibitem{crick_astonishingHyp_1994}
F.~Crick and J.~Clark, ``The astonishing hypothesis,'' {\em Journal of
  Consciousness Studies}, vol.~1, no.~1, pp.~10--16, 1994.

\bibitem{kim_MakeSenseEmerg_1999}
J.~Kim, ``Making sense of emergence,'' {\em Philosophical studies}, vol.~95,
  no.~1, pp.~3--36, 1999.

\bibitem{brenner_SeqsConseqs_2010}
S.~Brenner, ``Sequences and consequences,'' {\em Philosophical Transactions of
  the Royal Society B: Biological Sciences}, vol.~365, no.~1537, pp.~207--212,
  2010.

\bibitem{emmeche_explaining_1997}
C.~Emmeche, S.~K{\o}ppe, and F.~Stjernfelt, ``Explaining emergence: towards an
  ontology of levels,'' {\em Journal for General Philosophy of Science},
  vol.~28, pp.~83--117, Jan. 1997.

\bibitem{de_haan_how_2006}
J.~de~Haan, ``How emergence arises,'' {\em Ecological Complexity}, vol.~3,
  pp.~293--301, Dec. 2006.

\bibitem{bedau_WeakEmergence_1997}
M.~A. Bedau, ``Weak emergence,'' {\em No{\^u}s}, vol.~31, no.~s11,
  pp.~375--399, 1997.

\bibitem{crutchfield_evolRevol_2003}
J.~P. Crutchfield, ``When evolution is revolution-origins of innovation,'' {\em
  Evolutionary Dynamics: Exploring the Interplay of Selection, Neutrality,
  Accident, and Function}, pp.~101--134, 2003.

\bibitem{bedau_downward_2002}
M.~Bedau, ``Downward causation and the autonomy of weak emergence,'' {\em
  Principia}, vol.~6, no.~1, p.~5, 2002.

\bibitem{holland_ChaosOrder_2000}
J.~H. Holland, {\em Emergence: From chaos to order}.
\newblock OUP Oxford, 2000.

\bibitem{huneman_DynamEmergence_2008}
P.~Huneman and P.~Humphreys, ``Dynamical emergence and computation: An
  introduction,'' {\em Minds and Machines}, vol.~18, no.~4, pp.~425--430, 2008.

\bibitem{huneman_LessonComputEmerg_2012}
P.~Huneman, ``Determinism, predictability and open-ended evolution: lessons
  from computational emergence,'' {\em Synthese}, vol.~185, no.~2,
  pp.~195--214, 2012.

\bibitem{bar-yam_mathematical_2004}
Y.~Bar-Yam, ``A mathematical theory of strong emergence using multiscale
  variety,'' {\em Complexity}, vol.~9, no.~6, pp.~15--24, 2004.

\bibitem{bich_Synthese_2012}
L.~Bich, ``Complex emergence and the living organization: an epistemological
  framework for biology,'' {\em Synthese}, vol.~185, no.~2, pp.~215--232, 2012.

\bibitem{ryan_emergence_2007}
A.~J. Ryan, ``Emergence is coupled to scope, not level,'' {\em Complexity},
  vol.~13, no.~2, pp.~67--77, 2007.

\bibitem{seth_measuring_2010}
A.~K. Seth, ``Measuring autonomy and emergence via granger causality,'' {\em
  Artificial Life}, vol.~16, pp.~179--196, Jan. 2010.

\bibitem{machta_paramCompress_2013}
B.~B. Machta, R.~Chachra, M.~K. Transtrum, and J.~P. Sethna, ``Parameter space
  compression underlies emergent theories and predictive models,'' {\em
  Science}, vol.~342, no.~6158, pp.~604--607, 2013.

\bibitem{sambin_SomePointsFormalTop_2003}
G.~Sambin, ``Some points in formal topology,'' {\em Theoretical computer
  science}, vol.~305, no.~1, pp.~347--408, 2003.

\bibitem{boniolo_vagueness_2008}
G.~Boniolo and S.~Valentini, ``Vagueness, kant and topology: a study of formal
  epistemology,'' {\em Journal of Philosophical Logic}, vol.~37, no.~2,
  pp.~141--168, 2008.

\bibitem{alley_OrganismEnvEpistem_1985}
T.~R. Alley, ``Organism-environment mutuality epistemics, and the concept of an
  ecological niche,'' {\em Synthese}, vol.~65, no.~3, pp.~411--444, 1985.

\bibitem{maturana_autopoiesisAndCognition_1991}
H.~R. Maturana and F.~J. Varela, {\em Autopoiesis and cognition: The
  realization of the living}, vol.~42.
\newblock Springer Science \& Business Media, 1991.

\bibitem{valentini_binaryPos_Unpublished}
S.~Valentini, ``The problem of completeness of formal topologies with a binary
  positivity predicate and their inductive generation,'' {\em Unpublished
  draft}.

\bibitem{anderson_MoreIsDiff_1972}
P.~W. Anderson {\em et~al.}, ``More is different,'' {\em Science}, vol.~177,
  no.~4047, pp.~393--396, 1972.

\bibitem{bohm_causality_1971}
D.~Bohm, {\em Causality and Chance in Modern Physics}.
\newblock University of Pennsylvania Press, Jan. 1971.

\bibitem{georgescu-roegen_entropy_1971}
N.~Georgescu-Roegen, {\em The Entropy Law and the Economic Process}.
\newblock Harvard University Press, first~ed., Jan. 1971.

\bibitem{stock_concepts-rels_2010}
W.~G. Stock, ``Concepts and semantic relations in information science,'' {\em
  Journal of the American Society for Information Science and Technology},
  vol.~61, no.~10, pp.~1951--1969, 2010.

\bibitem{ives_stabilityRev_2007}
A.~R. Ives and S.~R. Carpenter, ``Stability and diversity of ecosystems,'' {\em
  Science}, vol.~317, no.~5834, pp.~58--62, 2007.

\bibitem{bunge_emergence_2003}
M.~Bunge, {\em Emergence and Convergence: Qualitative Novelty and the Unity of
  Knowledge}.
\newblock University of Toronto Press, 2003.

\bibitem{strunz_resilience-vagueness_2012}
S.~Strunz, ``Is conceptual vagueness an asset? arguments from philosophy of
  science applied to the concept of resilience,'' {\em Ecological Economics},
  vol.~76, pp.~112--118, 2012.

\bibitem{hooker_Constraints_2013}
C.~Hooker, ``On the import of constraints in complex dynamical systems,'' {\em
  Foundations of Science}, vol.~18, no.~4, pp.~757--780, 2013.

\bibitem{auyang_CsystTheories_1999}
S.~Y. Auyang, {\em Foundations of complex-system theories: in economics,
  evolutionary biology, and statistical physics}.
\newblock Cambridge University Press, 1999.

\bibitem{bialek_flocking_2012}
W.~Bialek, A.~Cavagna, I.~Giardina, T.~Mora, E.~Silvestri, M.~Viale, and A.~M.
  Walczak, ``Statistical mechanics for natural flocks of birds,'' {\em
  Proceedings of the National Academy of Sciences}, vol.~109, no.~13,
  pp.~4786--4791, 2012.

\bibitem{bersini_Synthese_2012}
H.~Bersini, ``Emergent phenomena belong only to biology,'' {\em Synthese},
  vol.~185, no.~2, pp.~257--272, 2012.

\bibitem{gellmann_InfoMeasures_1996}
M.~Gell-Mann and S.~Lloyd, ``Information measures, effective complexity, and
  total information,'' {\em Complexity}, vol.~2, no.~1, pp.~44--52, 1996.

\bibitem{boschetti_intervention_2011}
F.~Boschetti, ``Causality, emergence, computation and unreasonable
  expectations,'' {\em Synthese}, vol.~181, no.~3, pp.~405--412, 2011.

\bibitem{bitbol2012downward}
M.~Bitbol, ``Downward causation without foundations,'' {\em Synthese},
  vol.~185, no.~2, pp.~233--255, 2012.

\bibitem{boniolo_objects_2012}
G.~Boniolo, S.~Valentini, {\em et~al.}, ``Objects: A study in kantian formal
  epistemology,'' {\em Notre Dame Journal of Formal Logic}, vol.~53, no.~4,
  pp.~457--478, 2012.

\bibitem{hoel-tononi_MacroBeatMicro_2013}
E.~P. Hoel, L.~Albantakis, and G.~Tononi, ``Quantifying causal emergence shows
  that macro can beat micro,'' {\em Proceedings of the National Academy of
  Sciences}, vol.~110, no.~49, pp.~19790--19795, 2013.

\bibitem{hegel_georg_2010}
G.~W.~F. Hegel and G.~D. Giovanni, {\em Georg Wilhelm Friedrich Hegel: The
  Science of Logic}.
\newblock Cambridge University Press, Aug. 2010.

\bibitem{legendre_numerical_2012}
P.~Legendre and L.~Legendre, {\em Numerical Ecology}.
\newblock Elsevier, July 2012.

\bibitem{altschul_BLAST_1990}
S.~F. Altschul, W.~Gish, W.~Miller, E.~W. Myers, and D.~J. Lipman, ``Basic
  local alignment search tool,'' {\em Journal of molecular biology}, vol.~215,
  no.~3, pp.~403--410, 1990.

\bibitem{van_to100CitedPapers}
R.~Van~Noorden, B.~Maher, R.~Nuzzo, {\em et~al.}, ``The top 100 papers,'' {\em
  Nature}, vol.~514, no.~7524, pp.~550--553, 2014.

\bibitem{hotelling_PCA_1933}
H.~Hotelling, ``Analysis of a complex of statistical variables into principal
  components.,'' {\em Journal of educational psychology}, vol.~24, no.~6,
  p.~417, 1933.

\bibitem{chandler_closure_2000}
J.~L. Chandler and G.~E. Van~de Vijver, {\em Closure: Emergent organizations
  and their dynamics}, vol.~401.
\newblock Annals of the New York Academy of Science, 2000.

\bibitem{cordero_MicrobEvol_2014}
O.~X. Cordero and M.~F. Polz, ``Explaining microbial genomic diversity in light
  of evolutionary ecology,'' {\em Nature Reviews Microbiology}, vol.~12, no.~4,
  pp.~263--273, 2014.

\bibitem{pascual_Thesis_2015}
A.~Pascual-Garc{\'\i}a, {\em Emergent patterns in protein, microbial and
  mutualistic systems}.
\newblock PhD thesis, Universidad Aut{\'o}noma de Madrid, 2015.

\bibitem{mee-wang_PNAS_2014}
M.~T. Mee, J.~J. Collins, G.~M. Church, and H.~H. Wang, ``Syntrophic exchange
  in synthetic microbial communities,'' {\em Proceedings of the National
  Academy of Sciences}, vol.~111, no.~20, pp.~E2149--E2156, 2014.

\bibitem{mayr_ObjectSelection_1997}
E.~Mayr, ``The objects of selection,'' {\em Proceedings of the National Academy
  of Sciences}, vol.~94, no.~6, pp.~2091--2094, 1997.

\bibitem{Tikhonov_elife_2016}
M.~Tikhonov, ``Community-level cohesion without cooperation,'' {\em eLife},
  vol.~5, p.~e15747, jun 2016.

\end{thebibliography}
 
\end{document}